\documentclass[aps,pra,longbibliography,twocolumn,showpacs]{revtex4-1}
\usepackage{graphicx}
\usepackage{amssymb,amsfonts,amsmath}
\usepackage[colorlinks=true,citecolor=Cerulean,linkcolor=RubineRed,urlcolor=Cerulean]{hyperref}
\hypersetup{breaklinks=true}
\usepackage{graphicx}
\usepackage{color}
\usepackage[usenames,dvipsnames]{xcolor}
\usepackage{epstopdf}
\usepackage[normalem]{ulem}

\AtBeginDocument{%
    \newwrite\bibnotes
    \def\bibnotesext{Notes.bib}
    \immediate\openout\bibnotes=\jobname\bibnotesext
    \immediate\write\bibnotes{@CONTROL{REVTEX41Control}}
    \immediate\write\bibnotes{@CONTROL{%
    apsrev41Control,author="08",editor="1",pages="1",title="0",year="1"}}
     \if@filesw
     \immediate\write\@auxout{\string\citation{apsrev41Control}}%
    \fi
}%

\newcommand{\Heff}{H_{\kappa}}
\newcommand{\Hinf}{H_{\infty}}
\newcommand{\Hhw}{H^{\rm hw}_{\kappa}}

\newcommand{\im}{\mathrm{i}}

\newcommand{\Htm}{h_{\infty}}

\newcommand{\ce}{c_\theta}
\newcommand{\se}{s_\theta}

\begin{document}

\title{Protected cat states from kinetic driving of a boson gas}

\author{G.~Pieplow}
\affiliation{Departamento de F\'isica de Materiales, Universidad
Complutense de Madrid, E-28040 Madrid, Spain}

\author{C.~E.~Creffield}
\affiliation{Departamento de F\'isica de Materiales, Universidad
Complutense de Madrid, E-28040 Madrid, Spain}

\author{F.~Sols}
\affiliation{Departamento de F\'isica de Materiales, Universidad
Complutense de Madrid, E-28040 Madrid, Spain}

\date{\today}

\begin{abstract}
We investigate the behavior of a one-dimensional Bose-Hubbard gas in both a ring and a hard-wall box, whose kinetic energy is made to oscillate with zero time-average, which suppresses first-order particle hopping.
For intermediate and large driving amplitudes the system in the ring has similarities to the Richardson model, but with a peculiar type of pairing and an attractive interaction in momentum space. This analogy permits an understanding of some key features of the interacting boson problem.
The ground state is a
macroscopic quantum superposition, or cat state, of two many-body states collectively occupying opposite momentum eigenstates. Interactions give rise to a reduction (or modified
depletion) cloud that is common to both macroscopically distinct states. Symmetry
arguments permit a precise identification of the two orthonormal
macroscopic many-body branches which combine to yield the ground state. 
In the ring, the system is sensitive to variations of the effective flux but in such a way that the macroscopic superposition is preserved. 
We discuss other physical aspects that contribute to protect the cat-like nature of the ground state.

\end{abstract}
\maketitle

\section{Introduction}
The existence of macroscopic quantum superposition (MQS) states, or cat states, has long been one of the most counterintuitive predictions of quantum mechanics \cite{schrodinger1935gegenwartige}, as it is at odds with our daily perception of reality, where such states are not observed.
The collapse of the Schr\"odinger cat state into one of its branches prevents us from directly observing coherent superpositions of macroscopically distinct states. This collapse is understood to be induced by the decohering effect of a dissipative environment which, in particular, can be a measuring apparatus \cite{von1932mathematische,zurek1982,joos1985emergence,zurek1991physics}. Decoherence tends to act faster the larger the quantum system is \cite{joos1985emergence}.

Cat states may be viewed as a particular and extreme form of entanglement, in which various subsystems (at least some of them containing many particles) of a macroscopic system share their fate by highly correlating their response to specific questions asked in the form of a quantum measurement. 
So perhaps the most ubiquitous type of cat state is that formed in a
measurement process where a microscopic quantum system becomes strongly correlated with the measuring apparatus. However, this type of MQS state is almost universally very fragile because, due to the macroscopic and noisy nature of the apparatus, the compound system quickly decays into one of the possible branches. For a given experimental run, we always observe the macroscopic apparatus in a well-defined configuration, from which we infer the state in which the quantum system is left \footnote{This description applies literally only to the ideal measurement of a discrete quantum variable. We do not consider here other types of measurement in the present discussion.}.

Cat states are of great fundamental interest. For instance, they have been proposed to test macroscopic realism 
\cite{leggett2002testing,cavalcanti2008,opanchuk2016quantifying,knee2016strict,
reid2018bell}, which posits an observer-independent description of macroscopic objects whilst keeping the laws of quantum mechanics intact at the micro-level \cite{knee2016strict}.
In that context, the characterization of an MQS state is especially relevant. 
Cat states also promise to be relevant for practical applications such as atom 
interferometry beyond the Heisenberg limit \cite{kurn2004}, precision measurements \cite{bollinger1996}, 
quantum information tasks \cite{leibfried2005creation} and quantum metrology  \cite{giovannetti2004quantum}. 

Quite generally, cat states are difficult to realize because coupling to the environment quickly destroys the coherent superposition of macroscopic states \cite{wheeler2014quantum,joos1985emergence,zurek1991physics}. So the quest for the realization of MQS states is mostly about finding means to overcome decoherence and to probe the coherence between the macroscopically distinct states before it is lost. 

In this paper we present a system with the remarkable property of possessing a cat-type ground state which is unusually protected from decay into one of its branches. 
Following our previous work \cite{pieplow2018generation}, in which we introduced the concept of ``kinetic driving'', where a system's kinetic energy is made to oscillate periodically in time with zero time-average, we investigate the behavior of a one-dimensional interacting boson system, with both periodic and hard-wall boundary conditions. In each case the ground state is a cat state involving two branches in which two different nonzero momentum states are macroscopically occupied. 

In both the ring and the hard-wall case, the system's resilience against collapse shows important differences as compared with other setups hosting cat-like states. 
Most importantly, the circular boson superfluid preserves the MQS when subject to a spurious flux. Similarly, the gas between hard walls is fundamentally unaltered by a velocity drift due to gauge invariance.
Another important feature is that the two condensates share the depletion cloud (here labeled reduction cloud), which provides a degree of protection against atom losses. Other aspects of resilience are discussed later in the paper.

Thus our goal here is to study the detailed properties of the cat-like ground state for a kinetically-driven boson system in both a ring and a box, where atoms satisfy periodic and hard-wall boundary conditions, respectively.

This paper is arranged as follows. Section \ref{overview} is devoted to a review of the experimental literature on cat states and an overview of representative theoretical papers.
In section \ref{kinetic-driving} we briefly reprise the main results of kinetic driving found in Ref. \cite{pieplow2018generation}. In section \ref{two-mode-model} we explain how seemingly natural attempts to find a guiding simple picture fail. Notably, the naive picture of a ground state consisting of a coherent superposition of two opposite-momentum condensates each traveling with its own depletion cloud yields nonsensical results. Section \ref{large-kappa-toy-model} is devoted to an illuminating toy model that correctly describes the strong driving limit and qualitatively applies to a wider range of driving amplitudes. The resulting model Hamiltonian contains a pairing interaction which is a particular case of the general Richardson-Gaudin pairing model studied in the context of nuclear physics and small grain superconductivity \cite{richardson1965exact,dukelsky2004colloquium}. Besides that pairing we additionally have an attractive interaction in momentum space that favors the macroscopic occupation of a single momentum mode \cite{Heimsoth2012}.
We find that the depletion cloud of the conventional, undriven Bose-Hubbard (BH) model, with its well-known pairing structure, is replaced here by a reduction cloud based on a novel type of pairing and shared by the two condensate branches. 

In section \ref{many-body-plane-waves} we investigate the many-body structure of the cat branches in the ring case. We find that the they are considerably more complex than the exclusive occupation of a nonzero momentum state. The collective ground state involves many momentum Fock configurations and there is no clear criterion to assign each configuration to a given branch. Indeed some momentum configurations contribute to both branches. We use a symmetry argument to sharply define the two orthonormal branches of the cat-like ground state. In section \ref{kinetic-hard-walls}  we perform a similar study for the hard-wall box, suitably switching between a stationary-wave and a truncated plane-wave representation. This section concludes with a brief discussion of the case of harmonic confinement, whose ground state shows a similar MQS state structure, which underlines the robustness of the physical effects unveiled here.

The atypical character of effective Hamiltonian yields an unconventional particle current operator that, in both the ring and the box scenarios, yields a vanishing expectation value (for zero effective flux) for the two stationary states and cat branches forming the ground doublet. In a time-of-flight experiment, however, the behavior of the two cat branches would be markedly different.

Section \ref{measures-of-cattiness} is devoted to a discussion of how different figures of merit help to measure the quality of the cat state. The discussion includes a brief review of the various ``cattiness'' measures that have been proposed in the literature, and introduces a new figure of merit. 
In section \ref{state-preparation} we investigate how realistic driving protocols enable the realization of the cat-like ground state. We find that the ground state is easier to reach in the ring case.
Finally, in section \ref{resilience} we discuss why this system is more resilient to collapse than most cat-state proposals. In particular, we consider the possibility of decay due to spurious rotations or velocity drifts, atom losses and diagonal impurities. We also contemplate thermal excitations and coupling to an environment as possible causes of collapse.

After a concluding section, the paper is complemented by Appendices dealing with a variational calculation and a cat measure estimate.

\section{Cat states}
\label{overview}

A cat state can be generally defined as the coherent superposition of two or a few macroscopically distinct states. 
Although the precise definition of `macroscopic' and `distinct' is by itself a subject of debate (as briefly reviewed in section \ref{measures-of-cattiness} for two branches), there is a considerable variety of physical systems for which cat states have been proposed and even realized.

Continuous superpositions of macroscopically distinct states (such as those appearing for double condensates \cite{sols1994,greiner2002}) may be considered MQS states, but usually are not referred to as cat states. The collapse of such states involves the measurement of a continuous macroscopic variable, which requires a specific discussion \cite{zapata2003} that we shall not address here. A double condensate system may evolve alternating between cat states and continuous MQS states \cite{greiner2002}.

Early experiments prepared atoms in a correlated superposition of spatially separated \cite{monroe1996schrodinger} or internal \cite{brune1996} states. Evidence of MQS states was also found in persistent current states in superconducting quantum interference devices (SQUIDs) \cite{van2000ch,friedman2000quantum} following earlier experimental work on macroscopic quantum behavior in those systems \cite{esteve1989observation,silvestrini1997,rouse1995}, or in related Cooper pair boxes \cite{nakamura1999coherent}. 

Diffraction, a characteristic quantum wave phenomenon, was observed in molecules as heavy as C$_{60}$ \cite{arndt1999wave} and even in larger molecules \cite{nairz2003quantum,eibenberger2013matter,gerlich2011quantum}. 
In the context of magnetism, coherent transitions between macroscopically distinct states involving different magnetic orientations of large molecules were observed \cite{PhysRevLett.68.3092,chudnovsky1993macroscopic,friedman1996}. 

Entanglement of up to four particles in trapped ion systems was detected in Ref. \cite{sackett2000experimental}. In similar systems, the number of entangled particles has later been increased to six \cite{leibfried2005creation} and fourteen \cite{monz2011}.
Photonic systems have been tailored to yield small cat states \cite{ourjoumtsev2006generating,ourjoumtsev2007generation} in the optical and also in the microwave range \cite{kirchmair2013,wang2016schrodinger}. 
In the optical range these states have been increased to involve hundreds \cite{bruno2013displacement} 
or even hundreds of millions \cite{lvovsky2013observation} of photons. 
In the microwave range cats of up to hundreds of photons have been produced \cite{vlastakis2013deterministically}.  
Cat states have also been realized in micromechanical oscillators \cite{o2010quantum}, sometimes 
involving optical levitation of submicron particles \cite{kiesel2013cavity}.
In the context of cold bosonic atoms, a double-well spinor cat was realized \cite{haycock2000}. 

On the theory side, early studies focused on superconducting devices \cite{leggetgarg1985,leggett2002testing}, including the coupling to a micromechanical resonator \cite{armour2002}. Other optomechanical oscillators have been proposed \cite{bose1999, marshall2003,shen2015nonlinear,liao2016}. It has been argued that even a mirror in a cavity can be prepared in a cat state \cite{bose1997preparation}.
Optical systems are also good candidates for MQS based on the superposition of coherent states with very different amplitudes \cite{stoler1986}. With the assistance of cavity collisions, few-atom systems can exhibit cat states of the Greenberger-Horne-Zeilinger (GHZ) type \cite{zheng2001}.

But perhaps the physical system that has collected the largest group of cat state proposals is the atom Bose-Einstein condensate 
\cite{zoller1998,ruos1998,dalvit200,kohler2001,polko2003,buonsante2005,dunningham2006,moore2006,korsbakken2007measurement,hallwood2006macroscopic,hallwood2007barriers,anamaria2007,nunnenkamp2008,carr2010,hallwood2010}, 
which is not surprising given its high tunability. 
To classify the various proposals, we may pay attention to the spatial distribution of the macroscopically distinct states, and the observable whose expectation value is different for the two branches. The most common choice is that of two (or a few) spatially separated atom clouds with the cat branches differing in the position of those clouds 
\cite{zoller1998,ruos1998,polko2003,moore2006,carr2010,haigh2010demonstrating}. Cat states may also exist with a similar spatial distribution but with the relative phase playing the role of the macroscopic observable distinguishing the two branches
\cite{kohler2001,dunningham2006,hallwood2010}. 
The branches may occupy the same region of space but populating modes
which differ in the density profile \cite{fischer2009, fischer2015}.
Another often-studied cat setup is that in which the two branches extend over the same ring but display different phase profiles, thus yielding distinct superfluid currents 
\cite{hallwood2006macroscopic,hallwood2007barriers,anamaria2007,nunnenkamp2008}.
For quantum gases extended over a ring, but described by
the attractive BH model, one may have all the atoms concentrated in a given site but hopping coherently between different sites in a highly correlated manner \cite{buonsante2005}. 
The atom spin can also be the observable distinguishing the two cat branches \cite{hoyip2000}, sometimes correlated with the condensate position \cite{dalvit200}.

We may also compare the various guiding principles that have been proposed to reach a cat state once it has been identified. When the MQS state is the ground state of a given system, simple cooling to that ground state
is very difficult because of the tiny energy splitting that separates the ground and first excited states \cite{daki2017}. Other approaches must thus be followed which can be roughly classified in three main categories: (i) projective measurement, (ii) dynamic evolution, and (iii) adiabatic preparation. 

Examples in the first category are provided in Refs.~\cite{dunningham2006, mazets2008creation}, where the creation of  macroscopic superpositions of atoms by some form of measurement on a Bose Einstein condensate was proposed. The setup  usually involves a system in a continuous superposition of macroscopically different states that is subject to a limited measurement unable to resolve the difference between between two distinct eigenvalues of the measured observable. A similar situation was found in Ref. \cite{kohler2001} for the relative phase of two interfering condensates.

The proposals in Refs.~\cite{stoler1986,kurn2004,moore2006,carr2010,nunnenkamp2008} can be framed in the second category. Dynamic preparation means that an initial state is prepared and made to evolve in such a way that sometime during the evolution the system is known to be in a cat state. 
A variant of the proposal of Ref.~\cite{stoler1986} was realized experimentally in a photonic system \cite{kirchmair2013}. 
The preparation of the initial state can be achieved by a sudden parameter change, for instance, the flip of the sign of the interaction in a bosonic atom gas \cite{moore2006}, or a sudden change of the tunneling phase \cite{nunnenkamp2008}.

Adiabatic preparation (third method) has also been proposed to produce cat states \cite{moore2006,zoller1998}. The authors of Ref. \cite{zoller1998} considered cooling to a ground state that is not cat-like but which can be adiabatically transformed into a cat state. 
Adiabatic preparation was implemented in the experiment of Ref.~\cite{friedman2000quantum}. There, a superposition of two macroscopically different flux states was created by slowly driving a SQUID towards a level anti-crossing. For superfluid systems, a general discussion of the difficulties to prepare MQS states in a rotating ring is presented in Ref. \cite{hallwood2007barriers}.

Finally, we may refer to the mechanisms that destroy the MQS state. The main cause of the fast decay of a Schr\"odinger cat state is the collapse into one of its branches due to decoherence \cite{zurek1982}. The detailed mechanism is very sensitive to the particular physical realization but, as already stated, it is a general trend that the larger and more different the superposed branches are, the faster the collapse is.
Particle losses  are known to be an important source of decoherence for MQS states in ultracold atoms \cite{zoller1998} or photonic systems \cite{Glancy2008}.
Interaction of the condensate with the depletion cloud plays an important role in destroying the MQS of boson condensates in two immiscible internal states \cite{dalvit200}.
The authors of \cite{moore2006} identify the trapping lasers and the thermal cloud of a trapped condensate as the main cause of the loss of quantum coherence. 
Interestingly, some physical effects tend to inhibit decoherence. For instance, inhomogeneities in a ring \cite{nunnenkamp2008} or interactions in a double well \cite{carr2010} can considerably delay the decay of the cat state.
A discussion of the robustness of the more general concept of fragmented states, of which cat states are a subset, can be found in Ref. \cite{uwe2013}

The unusual resilience to collapse is precisely one of the most interesting features of the setups considered in this paper. Section \ref{resilience} contains a discussion of the possible mechanisms that may induce the decay of the cat state.

\section{Kinetic driving in the ring}
\label{kinetic-driving}

In a previous article \cite{pieplow2018generation} we considered a kinetically driven one dimensional Bose-Hubbard (BH) model with $L$ sites and periodic boundary conditions, i.e. a ring. We chose a specific time periodic hopping amplitude (hence kinetic driving), which made our starting point the Hamiltonian
\begin{align}
\begin{aligned}
\mathcal{H}(t) = -J\cos(\omega t) \sum_{x=0}^{L-1}&( a_x^\dagger a_{x+1} + a_{x+1}^\dagger a_x)\\
 &+ \frac{U}{2} \sum_{x=0}^{L-1} n_x(n_x-1)~,
\label{eq:driven_BH}
\end{aligned}
\end{align}
where $a_x,a_x^\dagger$ are the usual bosonic annihilation(creation) operators and $n_x = a_x^\dagger a_x$ is the number operator. 
The Hubbard interaction energy is given by $U>0$, and $J\cos(\omega t)$ is the time periodic tunneling amplitude between neighboring sites.  We set $\hbar = 1$. In \cite{pieplow2018generation} we derived an effective {\em static} model in the limit of high frequency driving. The interested reader is referred to the appendix of \cite{pieplow2018generation} for details of the derivation.    
In the (quasi-)momentum representation, the effective Hamiltonian for the driven system \eqref{eq:driven_BH} is 
\begin{align}
\begin{aligned}
\Heff = &\frac{U}{2L}\sum^{L-1}_{l,m,n,p=0} \mathcal{J}_0 [ 2  \kappa F(k_l,k_m,k_n,k_p) ] \\ 
&\times 
a_{k_p}^\dagger a_{k_n}^\dagger  a_{k_m} a_{k_l} \,
\delta_{k_l+k_m,k_n+k_p}
\, ,
\end{aligned}
\label{eq:Heff}
\end{align}
where 
\begin{equation}
F(k_l,k_m,k_n,k_p)\equiv \cos(k_l)+\cos(k_m)-\cos(k_n)-\cos(k_p) \, ,
\label{def-F}
\end{equation}
the Kronecker delta expresses momentum conservation mod $2\pi$, $\mathcal{J}_0$ is the zeroth-order Bessel function, and the momenta are given by  
\begin{equation}
k_p \equiv 2\pi p/L \, ,
\label{define-kp}
\end{equation}
with $p$ taking $L$ integer values from 0 to $L-1$,
and $\kappa \equiv J/\omega$ is the driving parameter. In the ring $k_p$ is physically equivalent to $k_p + 2\pi n$ with $n \in \mathbb{Z}$. 
The properties of the system are determined by $\kappa$ together with the system size $L$ and the total particle number $N$. We choose $N$ to be even. The interaction $U$ becomes a global scaling factor and is the only remaining energy scale.  
The transformation relating the momentum and position representations is 
\begin{equation}
a_{k_p} = \frac{1}{\sqrt{L}}\sum_{x=0}^{L-1} e^{\im k_p x}a_{x}\,,\quad a_{x} = \frac{1}{\sqrt{L}}\sum_{p=0}^{L-1} e^{-\im k_p x}a_{k_p}\, .
\label{eq:pw_expansion}
\end{equation}
Next we briefly recapitulate some relevant results obtained in \cite{pieplow2018generation}.  
We found that the effective Hamiltonian \eqref{eq:Heff} has a quantum phase transition of the Kosterlitz-Thouless type.
Starting from a Mott-insulating state at $\kappa = 0$ the system undergoes a transition at $\kappa \simeq 0.48$ into a fragmented superfluid state, which is (quasi-)condensed into the momentum eigenstates of nonzero momenta $\pm \pi/2$. 
These momentum eigenstates have nonzero group velocity.
The condensation is signaled by both the onset of clearly separated peaks in the momentum density $\langle n_k \rangle$ at $k =\pm \pi/2$ 
\footnote{We adopt the convention of using the subindex $p$ (or equivalent) for the momentum $k_p$ [as in Eq. \eqref{eq:pw_expansion}] only when deemed necessary; otherwise, we plainly refer to $k$ or $k'$ with their sum understood as in the second Eq. \eqref{eq:pw_expansion}.}, and the closing of the Mott gap. Additional confirmation is obtained by measuring the Luttinger liquid parameter, $K_b$, which takes the universal value $K_b = 1/2$ at the Kosterlitz-Thouless transition point.

Further information can be extracted from the two-particle momentum density. We found that the momentum-momentum correlation $\langle n_{k} n_{k'}\rangle$ develops well separated peaks at $(k,k')=\pm (\pi/2, \pi/2)$. Remarkably, the correlation 
is much smaller at $(\pi/2,-\pi/2)$, which leads us to conclude that the ground state is a Schr\"odinger cat-like superposition of two macroscopically occupied momentum eigenstates, rather than something similar to a mere product state of $N/2$ particles in $\pi/2$, and $N/2$ in $-\pi/2$.

In Ref. \cite{pieplow2018generation} we provided some semiquantitative arguments to understand why the $\pm \pi/2$ states are macroscopically occupied. In this paper we present additional considerations that give us a deeper understanding of the special role played here by the momenta $\pm \pi/2$. We may already point out that, if we take the expectation value of \eqref{eq:Heff} in a many-body state consisting of all particles occupying the $k$ momentum eigenstate, the resulting energy is independent of $k$.
This is a first clear indication that interactions play a crucial role in determining the structure of the ground state. 

In the standard (undriven) Bose-Hubbard model the occupation of the zero momentum state is favored because it minimizes the kinetic energy invested in the coherent hopping between sites. Deep in the superfluid regime the interactions may only represent an energetically small perturbation to this state, but one that is essential to achieve superfluidity. Here the situation is more involved because $\Heff$ only consists of interactions, and does not permit single-particle nearest-neighbor hopping.

At this point we may also note some important symmetries. One can verify that $\Heff$ remains invariant under the transformations $k \rightarrow -k$ (time inversion) and $k \rightarrow \pm \pi + k$. This highlights the importance of the momentum eigenstates with momentum eigenvalues $0,\pi$ and $\pm \pi/2$, since the unordered pairs $(0,\pi)$ and $(\pi/2,-\pi/2)$ remain invariant under the mentioned symmetry transformations. Thus the two components of each of those pairs must play symmetrical roles. In particular, if the ground state wave function contains a term where $\pi/2$ is macroscopically occupied, then there must be another term where $-\pi/2$ has exactly that occupation. 

Later in the article (see Section \ref{large-kappa-toy-model} and Appendix \ref{app:var}) we will further argue why the macroscopic occupation of momenta $\pm \pi/2$ can be expected to be energetically favored.

\section{Two-mode model and failure of the Bogoliubov approximation}
\label{two-mode-model}

One can gain more analytical insight by assuming the presence of only two modes $a\equiv a_{-\pi/2}$, $b\equiv a_{\pi/2}$. Then the Hamiltonian \eqref{eq:Heff} becomes:
\begin{align}
\begin{aligned}
H_{\rm 2LS} = &\frac{U}{2 L}\left(
2 N^2-N-
n_{a}^2-n_{b}^2
\right.
\\
&
\left.+{a^\dagger}
^2 b^2 + {b^\dagger}^2a^2\right)~,
\label{2LS-Hamiltonian}
\end{aligned}
\end{align}
where we have used $n_a+n_b=N$, and the subindex 2LS stands for two-level system.

The above Hamiltonian can be diagonalized with the canonical transformation, $c = \frac{1}{\sqrt{2}}(a+b)~,d = \frac{1}{\sqrt{2}}(a-b)$, which leads to 
\begin{equation}
H_{\rm 2LS} = \frac{U}{2L}(2N^2- 2N - 4n_c n_d)~,
\end{equation}
the normalized ground state being 
\begin{equation}
|\Psi\rangle = \frac{1}{(N/2)!}\frac{1}{2^{N/2}}({a^\dagger}^2-{b^\dagger}^2)^{N/2}|{\rm vac}\rangle \label{eq:two_mode_gs}
\end{equation}
This state does not have the cat-like momentum-momentum correlations that we observe numerically, and which ideally would be represented by a many-body state of the type
\begin{equation}
(2N!)^{-1/2}[(a^{\dagger})^N \pm (b^{\dagger})^N]|\rm vac \rangle
\label{ideal-cat-state}
\end{equation}

Actually, it can be shown that in the large-$N$ limit \eqref{eq:two_mode_gs} satisfies $\langle n_{a} n_{b}\rangle/\langle n_{a} n_{a}\rangle \rightarrow 1/3$, which is much larger than the numerical result $\langle n_{a} n_{b}\rangle/\langle n_{a} n_{a}\rangle \ll 1$, while the same ratio is exactly zero for the state \eqref{ideal-cat-state}.
The reason for this discrepancy lies in the coefficients accompanying the ``center" configuration $|(N/2)_{a}, (N/2)_{b}\rangle$ when the binomial expansion is implemented in Eq. \eqref{eq:two_mode_gs} 
\footnote{We introduce the notation \unexpanded{$|M_a \rangle \equiv (M!)^{-1/2}(a^\dagger)^M |{\rm vac}\rangle$} whereby \unexpanded{$|M_a\rangle$} is a state where $M$ particles occupy mode $a$.}.
Their weights are simply too large. The failure of the two-mode approximation to reproduce the correct correlations further highlights the need to include the other modes ($k \neq \pm \pi/2$) and the interactions that connect them. In the next section we address this question more precisely through a simplified description of the superfluid regime. 

We note that, once we have decided to focus on the macroscopic occupation of two modes, \eqref{eq:two_mode_gs} is the closest analog here to the Gross-Pitaevskii (GP) solution of the undriven Bose-Hubbard model, where only the occupation of one mode is considered. The GP approximation assumes a fully occupied one-atom mode of the type
\begin{equation}
(N!)^{-1/2} (a^{\dagger})^N|{\rm vac}\rangle \, .
\label{GP-state}
\end{equation}

\begin{figure}[t!]
	\center
	\includegraphics[width=.45\textwidth]{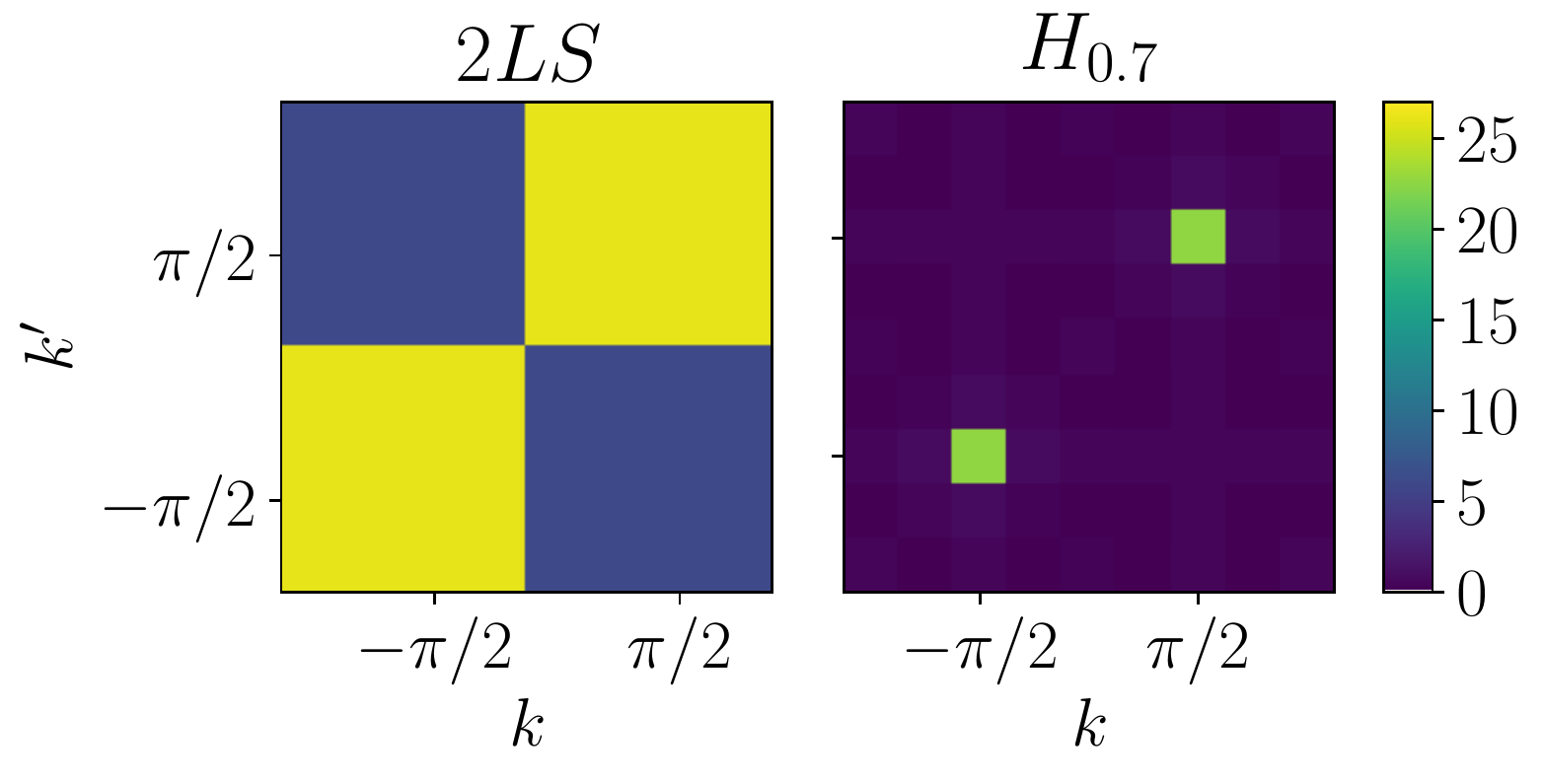}
	\caption{Momentum correlations for the two-mode approximation (left panel) and for the full effective Hamiltonian (right panel) at $\kappa=0.7$. The respective ratios $\langle n_{\pi/2} n_{\pi/2}\rangle/\langle n_{-\pi/2} n_{\pi/2}\rangle$ are $4.3$ and $45.5$, which means that the ground state of the two-mode approximation is much less cat-like than that of the full model.}
	\label{Fig:corr_comp}
\end{figure}

One might be tempted to follow the steps of the undriven Bose gas and replace the operators $a,b$ in \eqref{eq:Heff} and \eqref{2LS-Hamiltonian} by c-numbers. However, such an approximation is valid only when all relevant configurations have a large occupation of both modes $a$ and $b$, but we find numerically that this is not the case here. That approximation does apply to a state of the type \eqref{eq:two_mode_gs} [this is clearer when \eqref{eq:two_mode_gs} is written in terms of the $c,d$ operators] but, as already noted, the numerical results reveal that the system
is far from that configuration. Most importantly, the weight of configurations in which both modes ($\pm \pi/2$) are largely occupied is tiny. Specifically, we know from numerical inspection that $\langle n_{k} n_{k'}\rangle$ is very small away from $(k,k') = \pm (\pi/2,\pi/2)$, as can be appreciated in right panel of Fig. \ref{Fig:corr_comp}. 

On the other hand, the macroscopic occupation of $\pm \pi/2$ invites us to consider the picture of two condensates traveling in opposite directions, each one carrying its own depletion cloud. One might then perform a Bogoliubov calculation based on the macroscopic occupation of one of the two preferred momentum states, hoping that a meaningful physical picture will be obtained. However, such a simple-minded Bogoliubov calculation yields anomalous results, the main one being that all quasiparticles have zero energy. Thus it becomes clear that the equivalent of the depletion cloud for the fragmented, cat-like condensate has to be qualitatively different from the conventional case. We will refer to
the set of momentum-representation Fock states with at least one momentum $k\neq \pm \pi/2$ occupied as the 
{\em reduction cloud}, to clearly distinguish it from the conceptually simpler depletion cloud of the undriven Bose-Hubbard model. In the next section we study an approximate Hamiltonian that provides valuable insights on the structure of the reduction cloud.

\section{Large $\kappa$ limit. Toy model.}
\label{large-kappa-toy-model}

As we have just argued, we need to fundamentally depart from the Bogoliubov-de Gennes approach, and turn to other approximation schemes. 
In order to investigate the structure of the reduction cloud, we resort to a simplified Hamiltonian and study it in the superfluid regime. We will assume that, in Eq. \eqref{eq:Heff}, the only relevant scattering processes are those for which the argument of the Bessel function is zero. This approximation becomes exact in the the limit $\kappa \rightarrow \infty$.

\begin{figure}[t!]
\center
\includegraphics[width=.338\textwidth]{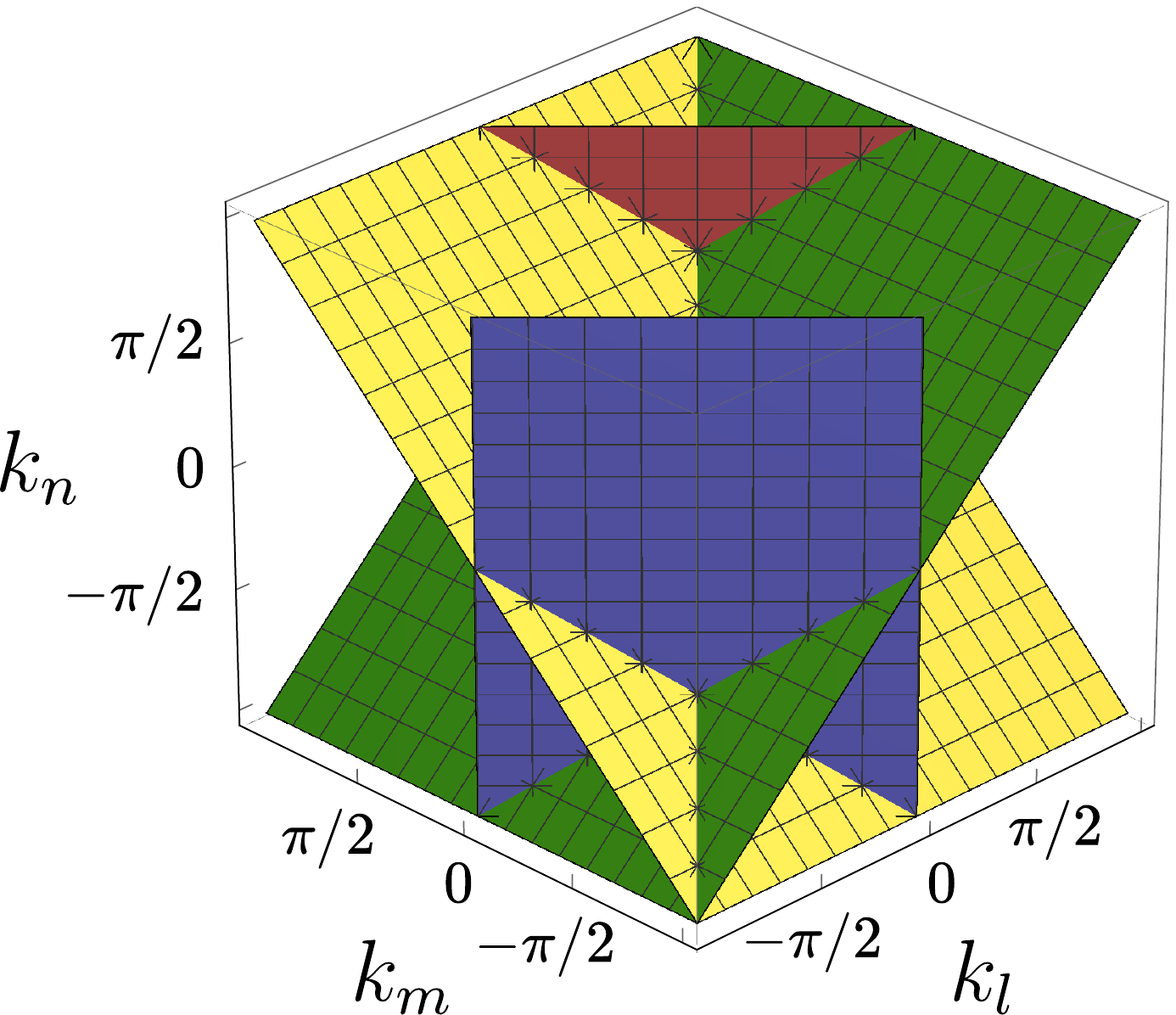}
\caption{The four planes (each plotted with a different color) containing the scattering events whose amplitude in $\Heff$ is independent of $\kappa$. We only show the first Brillouin zone in $k_l,k_m,k_n$-space.}
\label{Fig:planes}
\end{figure}

For those elementary processes, the matrix element is independent of $\kappa$. They satisfy the condition
\begin{equation}
F(k_l,k_m,k_n,k_p)=0 \, ,
\label{Fequalzero}
\end{equation}
i.e., $\mathcal{J}_0(2\kappa F) = 1$ in \eqref{eq:Heff}. By invoking momentum conservation, \eqref{Fequalzero} becomes
\begin{equation}
\cos(k_l)+\cos(k_m)-\cos(k_m+k_l-k_n)-\cos(k_n) = 0\, .
\label{eq:constraintcond}
\end{equation}
This is true provided any of the following four conditions is fulfilled:
\begin{equation}
k_l = k_n~, \quad k_m = k_n~, \quad k_l + k_m = \pm \pi~,
\label{three-conditions}
\end{equation}
the first two describing the absence of a collision.
These three equations can be neatly viewed as four planes in the first Brillouin zone of $(k_l,k_m,k_n)$-space.
This is illustrated in Fig. \ref{Fig:planes}, where the third Eq. \eqref{three-conditions} appears unfolded into two planes
plotted in the same color.
Any point on these four planes represents a scattering event that has amplitude $U/2L$. Since $\lim_{x\rightarrow 0}\mathcal{J}_0(x) = 1$, these are the amplitudes that dominate in the large-$\kappa$ limit. 
In that limit $\Heff$ effectively becomes
\begin{align}
\begin{aligned}
H_{\infty} = &\frac{U}{2L} \left( 2N^2-N-\sum_{k} n_{k}^2 \vphantom{\left[\sum_{\substack{k_m\\ k_p}}\right.}\right.\\ 
\hspace{1.5cm}+& \left.\sum_{\substack{k\neq k' \\ k+k'\neq \pi }}a^\dagger_{\pi-k}a^\dagger_{k}a_{\pi-k'}a_{k'}\right)\, .
\end{aligned}
\label{eq:h_constr}
\end{align}
Hereafter the sums over momenta $k,k'$ are understood in the sense of \eqref{define-kp} and \eqref{eq:pw_expansion}.

\begin{figure}[t!]
\center
\includegraphics[width = .45 \textwidth]{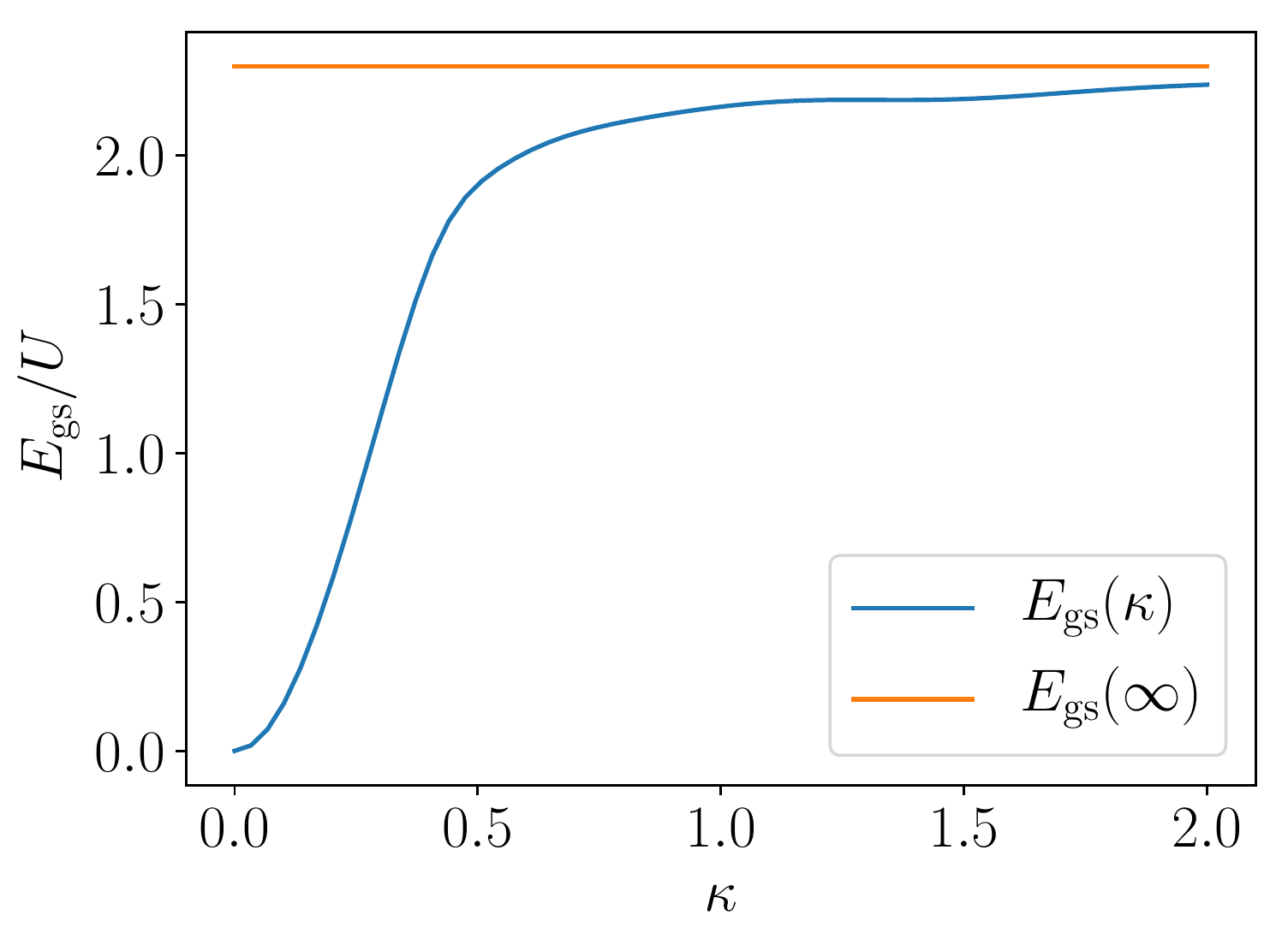}
\caption{Comparison of the ground state energy of $\Heff$ and $\Hinf$ for 8 particles on 8 sites. As expected, for increasing $\kappa$ the ground state energy of $\Heff$ approaches that of the asymptotic model. We have checked this convergence for multiple system sizes and fillings.}
\label{Fig:gr_comparison}
\end{figure}

\begin{figure}[t!]
\center
\includegraphics[width=.332\textwidth]{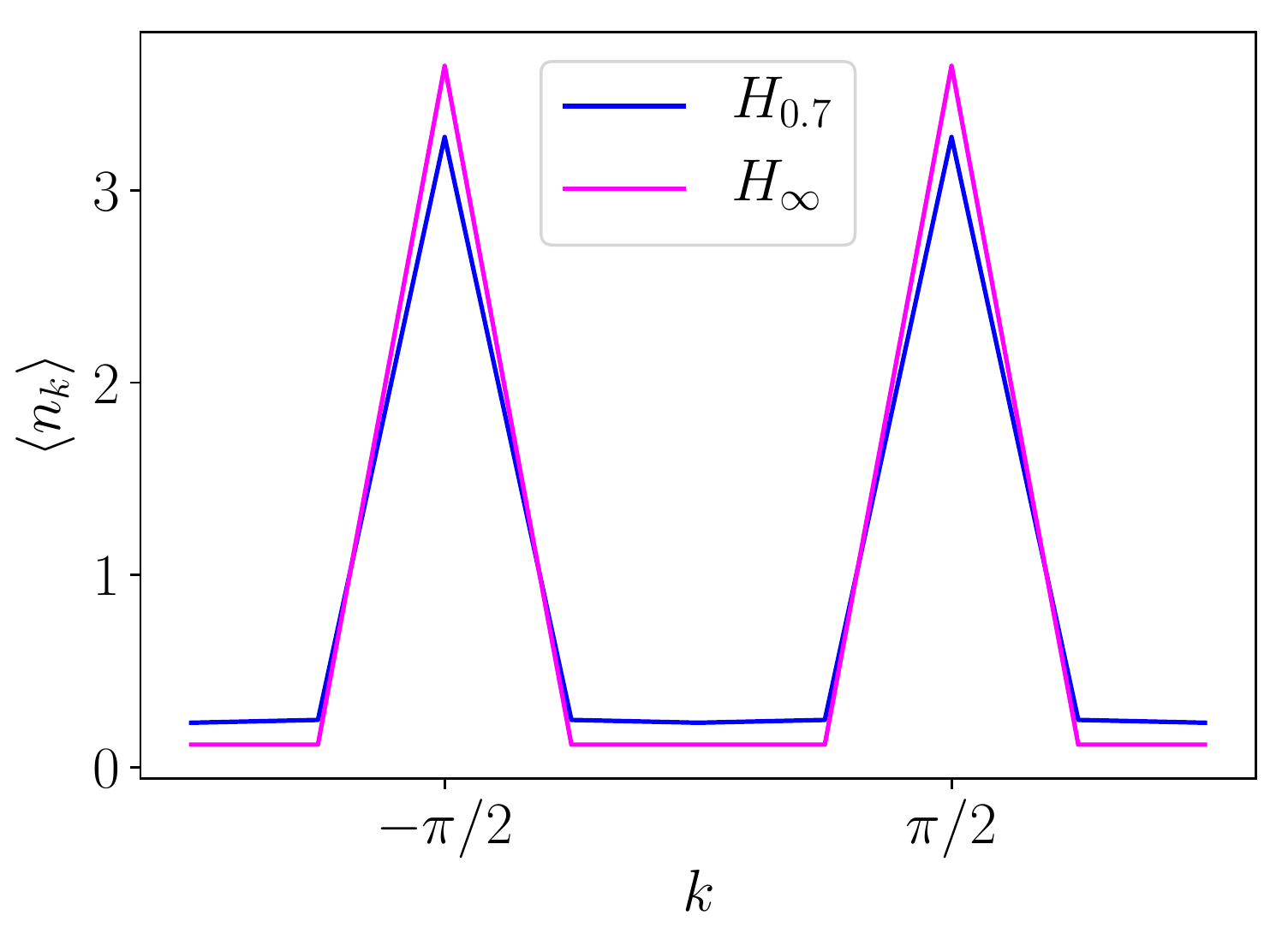}
\includegraphics[width=.45\textwidth]{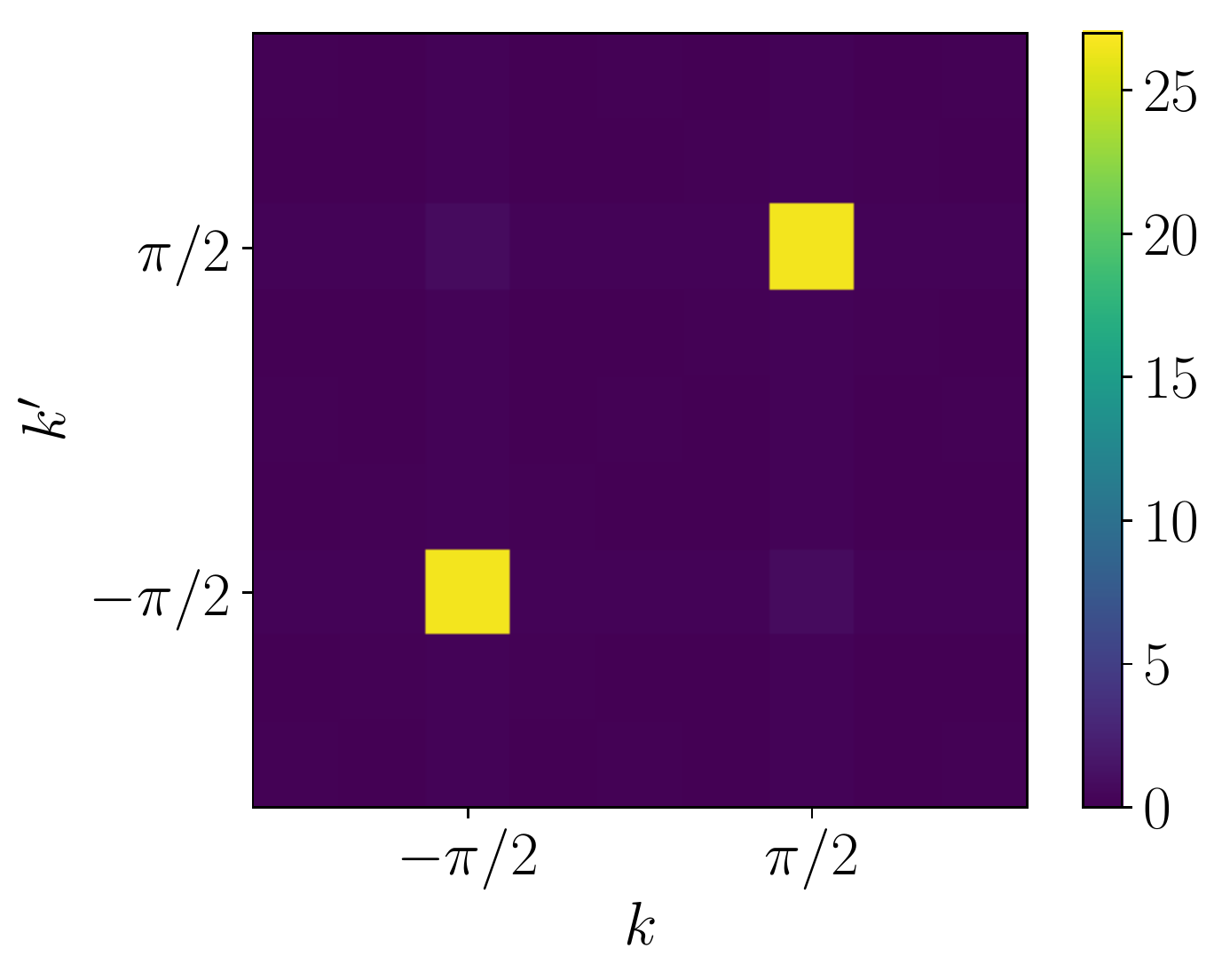}
\caption{Top: Momentum density $\langle n_k \rangle$ for the ground state of $H_{0.7}$ and $\Hinf$ for 8 particles on 8 sites. Both show clearly separated peaks around $k = \pm \pi/2$. Bottom: momentum-momentum correlations for the ground state of $\Hinf$. Again, they are clearly peaked around $\pm \pi/2$ and show negligible cross correlations, indicating a cat-like state rather than a product state. Note that the case $\kappa=0.7$ is plotted in right Fig. \ref{Fig:corr_comp}. 
}
\label{Fig:mom_dens_mom_corr} 
\end{figure}

Although this Hamiltonian is exact only for $\kappa \rightarrow \infty$, we will see that it adequately captures many of the 
properties of our system for moderately large values of $\kappa$. Figures \ref{Fig:gr_comparison} and \ref{Fig:mom_dens_mom_corr}a show that, even for $\kappa \alt 1 $, the ground state energy and the momentum density of $\Heff$ are very close
to those of $\Hinf$. Comparison of the right Fig. \ref{Fig:corr_comp} and the lower Fig. \ref{Fig:mom_dens_mom_corr} shows that the two-particle momentum densities are also very similar.
This justifies the detailed study of $\Hinf$ as a reference model for the more physical parameter regime of moderate $\kappa$. 
By removing the scaling factor $U/2L$, and the terms that are constant for a given $N$, we trim $\Hinf$ 
down to the simple toy model 
\begin{equation}
h_{\infty} = -\sum_{k} n_{k}^2 + \sum_{\substack{k \neq k' \\ k+k' \neq \pi} }a^\dagger_{\pi-k}a^\dagger_{k}a_{\pi-k'}a_{k'} \,
\label{eq:pairing}
\end{equation}
which adequately captures, for a broad range of $\kappa$ values ($\kappa \sim 0.5$ and larger), the properties of the true ground state in the superfluid sector.

\subsection{Connection to the Richardson model}

The toy model \eqref{eq:pairing} has interesting connections to other areas of physics. The second term shows a pairing-type interaction in which only collisions between particle pairs with total momentum $\pi$ are allowed. 

Pairing interactions (albeit typically between pairs of zero total momentum) appear in the theory of superconductivity and in nuclear physics contexts. This topic has a rich history in which many exactly solvable models have been developed and studied. An excellent review on this topic can be found in Ref. \cite{dukelsky2004colloquium}, where frequent references are made to the work of Richardson \cite{richardson1965exact}, who found a numerically exact solution for what has later become known as the Richardson model. The theory was rediscovered and successfully applied to mesososcopic superconducting metallic grains \cite{sierra2000exact, fazio2001}.
Since then a whole class of integrable so-called Richardson-Gaudin models have been found \cite{dukelsky2004colloquium}. Despite $\Htm$ not being included in that set of models, we can make use 
of many of the tools developed in \cite{richardson1965exact}, and the forthcoming analysis is an adaption of elements of Richardson's theory to our model.
 
The main difference between the Richardson model and $\Htm$ is the first term on the r.h.s. of Eq. 
\eqref{eq:pairing}, which, as noted in Ref. \cite{Heimsoth2012}, may be viewed as an attractive interaction in momentum space. 
A particle pair in our system is created by the operator $B_{k}^\dagger = a^\dagger_{\pi-k}a^\dagger_{k}$. 
Just as in Ref. \cite{richardson1965exact}, we can define the seniority operator 
\begin{equation}
\nu_k = |n_{k}-n_{\pi-k}|\, ,  
\end{equation} 
which counts the number of unpaired particles with momentum $k$ or $\pi - k$
\footnote{Actually in Ref. \cite{richardson1965exact} seniority was defined in a formally different way because it was initially applied to fermions.}.
We note that 
\begin{equation}
[\Htm,\nu_k]=0~,
\label{seniority-commutes}
\end{equation}
and define the seniority of an eigenstate of $\Htm$ as its eigenvalue with respect to the operator
\begin{equation}
\nu = \frac{1}{2}\sum_{k} \nu_{k}~,
\label{total-seniority}
\end{equation}
which measures the total number of unpaired particles, the factor $1/2$ being introduced to prevent double counting.

The conservation of seniority ($[\Htm,\nu]=0$) permits a helpful block-diagonalization of $\Htm$, since only configurations with the same seniority are connected by $\Htm$. 
In particular, states that do not contain any pairs ($\nu = N$) are eigenstates of the pairing term with eigenvalue $0$ and thus eigenstates of $\Htm$. The states with the lowest 
energy within the class of $\nu = N$ are those of the type $|N_{k\neq \pm \pi/2}\rangle$, their energy being $-N^2$, where
\begin{equation}
|N_{q}\rangle \equiv (N!)^{-1/2}(a_{q}^\dagger)^N|\rm vac \rangle
\label{def-Nq}
\end{equation}
is a state with $N$ particles in momentum $q$. 

An important sector in the block-diagonalized Hamiltonian is $\nu = 0$, i.e., the set of states that are only made of configurations where {\em all} particles are paired. We refer to them as fully paired states. 
Within this set of configurations, only $|N_{\pm \pi/2}\rangle$ involve the exclusive occupation of a single one-atom state, since $\pm \pi/2$ are the only two momenta that form a pair with themselves.
Similarly to $|N_{k\neq \pm \pi/2}\rangle$, they benefit the most from the ``on-site'' attractive interaction in momentum space. However, unlike $|N_{k\neq \pm \pi/2}\rangle$, $|N_{\pm \pi/2}\rangle$ are not eigenstates of the repulsive pairing term in \eqref{eq:pairing}, because of the contribution
\begin{equation}
\sum_{k \neq \pm \pi/2}
a^\dagger_{\pi-k}a^\dagger_{k}\left(a_{\pi/2}a_{\pi/2}+ a_{-\pi/2}a_{-\pi/2}\right )
+ {\rm H.c.}
\label{pi-2-k-different}
\end{equation}
there included. Intuitively one expects that states involving only $\pi/2$ or $-\pi/2$ or both will further lower their energy by mixing with configurations which include pairs
with $k \neq \pm \pi/2$. As announced in the previous section, we will refer to those states as the reduction cloud of the $\pm \pi/2$ condensates.

The main differences between the reduction cloud and the conventional depletion cloud are that (i) the pairs of the reduction cloud have total momentum $\pi$ (in contrast to zero for the depletion cloud) and, most importantly, (ii) the reduction cloud is shared by the two condensates. The reason for the second difference is clear from \eqref{pi-2-k-different}: the creation of a pair $(k, \pi - k)$ of total momentum $\pi$ (with $k \neq \pm \pi/2$) can borrow the momentum indistinguishably
from condensate pairs $(\pi/2, \pi/2)$ or $(-\pi/2, -\pi/2)$, which have the same total crystal momentum, namely, $\pi$.

Indeed it can be proven by a variational calculation that mixing 
$|N_{\pm \pi/2}\rangle$ with other fully paired (but containing at least one $k\neq \pm \pi/2$)
further lowers the energy beyond $-N^2$. The details of this can be found in Appendix \ref{app:var}. This result confirms the intuition that the two states $|N_{\pm \pi/2}\rangle$ benefit energetically from the mixing with the interaction-induced reduction cloud. 
More specifically, 
$|N_{\pm \pi/2}\rangle$ mixes with other fully paired states of the type $|(N-2)_{\pm \pi/2},1_{k},1_{\pi -k}\rangle$ with $k\neq \pm \pi/2$, which lowers the energy thanks to the pairing interactions in $\Heff$. By contrast, states of the type $|N_{k}\rangle$ with $k\neq \pm \pi/2$ do not have this possibility, since the pairing term does not allow them to mix with other configurations. As already noted, they are eigenstates of $\Heff$.

\begin{figure}[t!]
	\center
	\includegraphics[width=.45\textwidth]{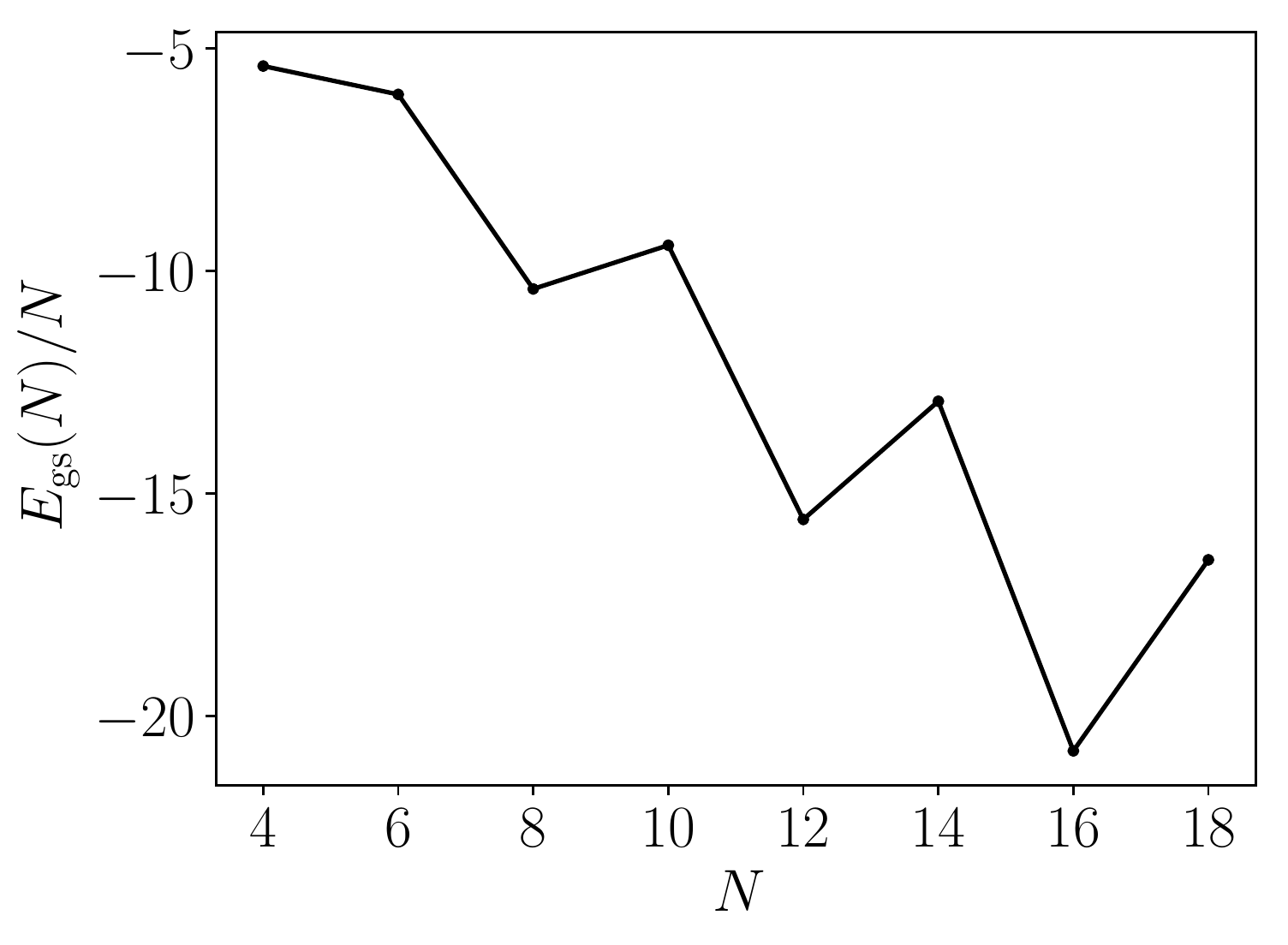}
	\caption{Energy per particle of the ground state of $\Htm$ for different system sizes with unit filling. $N=L$ is the number of sites in the ring. A system that does not include $\pm \pi/2$ in its set of momenta has a slightly higher energy per particle than an adjacent system of similar size that does contain it. The latter case corresponds to $N/4$ being an integer.}
	\label{Fig:gs_en_scaling}
\end{figure}

The block diagonalization greatly helps in reducing the numerical overhead when studying the ground state of the system. For $\nu = 0$ and unit filling, one only has to deal with
$\binom{N-1}{N/2}$ configurations instead of $\binom{2N-1}{N}$ \footnote{Recall that $N$ bosons distributed in $L$ sites yield $\binom{N+L-1}{N}$ distinct configurations.}. For 8 particles on 8 sites 
this already makes a big difference. Instead of $6435$ configurations, one only has to work with $70$. This allows the system size to be increased  up to 18 particles on 18 sites. 
In Fig. \ref{Fig:gs_en_scaling} we see an interesting scaling of the energy per particle when increasing the system size. The energy drops in a zig-zag pattern. This is connected to the fact that 
only when $\pm \pi/2$ are among the allowed momenta can we have fully paired states involving the occupation of just two modes, namely, $\pm \pi/2$.
This is possible only when $N/4$ is an integer.

\subsection{Ground state}

Inspection of $\Htm$ shows that all the momenta $k\neq \pm \pi/2$ play an equivalent role. This suggests that the ground state of $\Htm$ can be written as a fully paired state of the form:
\begin{widetext}
\begin{align}
\begin{aligned}
|\Psi_0\rangle = &\sum_{m=0}^{N/2} \sum_{l=0}^{(N-2m)/2}
C_{m,l}\left[ |(N-2m-2l)_{-\frac{\pi}{2}},(2l)_\frac{\pi}{2}\rangle
+ |(2l)_{-\frac{\pi}{2}}, (N-2m-2l)_{\frac{\pi}{2}}\rangle \right]
\\ &\hspace{3cm}\times\sum_{\sum n_{k} = m}
 Q_{\{n_k\}} \sum_P |P\{n_k\}\rangle \, ,
\label{GS-toy-model-general}
\end{aligned}
\end{align}
\end{widetext}
where the sum over $\{n_k\}$ runs over all possible occupation numbers $n_k$ of the pairs $(k,\pi -k)$, each one characterized by the momentum $k$ satisfying $0 \le k < \pi/2$, so that in each sequence $\{n_k\}$, a given number $n_k$ represents the two-mode state $|n_{k},n_{\pi-k}\rangle$ containing $n$ particles in momentum $k$ and $n$ particles in momentum $\pi-k$. We also sum over all possible permutations $P$ of the sequence of pair occupation numbers. Each permutation $P$ acting on $\{n_k\}$ yields the same number sequence but distributed throughout the set of state pairs (always with $k\neq \pm\pi/2$) in a different way.
The resulting pair configuration is represented by $|P\{n_k\}\rangle$ and must have the same weight as $|\{n_k\}\rangle$. 
Thus the ansatz \eqref{GS-toy-model-general} is expected to be exact within the toy model, something which we have confirmed numerically.

To further understand the above state we may write down the leading contributions to the ground state. If $Q_m$ is defined as $Q_{\{n_k\}}$ for the particular sequence in which $m$ particle pairs transferred to the reduction cloud are concentrated in one mode pair [as usual labeled $(k,\pi-k)$ with $k \neq\pm\pi/2$], then we can write 
\begin{widetext}
\begin{align}
\begin{aligned}
|\Psi_0\rangle = 
& C_{0,0}Q_{0}\Big(|N_{-\frac{\pi}{2}},0_\frac{\pi}{2}\rangle
+
|0_{-\frac{\pi}{2}},N_{\frac{\pi}{2}}\rangle\Big) 
+ 
C_{1,0}Q_{1}\Big(|(N-2)_{-\frac{\pi}{2}},0_\frac{\pi}{2}\rangle
+
|0_{-\frac{\pi}{2}},(N-2)_{\frac{\pi}{2}}\rangle \Big)\sum_{k\neq \pm \pi/2}|1_{k}, 1_{\pi-k}\rangle 
\\ &
+C_{0,1}Q_{0}\Big(|(N-2)_{-\frac{\pi}{2}},2_\frac{\pi}{2}\rangle
+
|2_{-\frac{\pi}{2}},(N-2)_{\frac{\pi}{2}}\rangle\Big) \\
&+ C_{1,0}Q_{2}\Big(|N-4)_{-\frac{\pi}{2}},0_\frac{\pi}{2}\rangle
+
|0_{-\frac{\pi}{2}},(N-4)_{\frac{\pi}{2}}\rangle\Big)\sum_{k\neq \pm \pi/2}|2_{k}, 2_{\pi-k}\rangle + \ldots \, ,
\end{aligned}
\label{eq:gs_toy_model}
\end{align}
\end{widetext}
where the terms are written in order of decreasing value of $|C_{m,l}Q_{m}|^2$, as obtained numerically for the case of 8 particles on 8 sites \footnote{Note that these decreasing coefficients are not necessarily correlated with the relative weight, within the many-body ground state, of a given type of configuration, since they are multiplying non-normalized many-body states.}.

The expansion \eqref{eq:gs_toy_model} has some interesting information. The first term is clearly identifiable as the ideal Schr\"odinger cat-like superposition of the two macroscopically occupied orbitals $\pm \pi/2$ (sometimes called a NOON state). The second term is the largest contribution to the reduction cloud. 
The third term represents an internal exchange of one pair between the two main configurations of the fragmented condensate, without intervention of the reduction cloud. The fourth term represents the exchange of four particles between the fragmented condensate and the reduction cloud all going to the same mode pair; and so on.

The many-body states \eqref{GS-toy-model-general} and \eqref{eq:gs_toy_model} show very clearly that the reduction cloud is shared by the two branches (or macroscopically distinct configurations) of the cat state. This fact definitely eliminates the naive picture of the cat state formed by two macroscopic branches each carrying its own depletion cloud. Some of these points are further discussed in Appendix \ref{app:var}.

As we have noted, when a particle pair of total momentum $\pi$ is created, its momentum can be equivalently viewed as coming from either the $\pi/2$ or the $-\pi/2$ condensates. In practice this means that, as we see in Eqs. \eqref{GS-toy-model-general} and \eqref{eq:gs_toy_model}, such a pair (with $k\neq\pm\pi/2$) factors out from a coherent, still cat-like superposition of the two different macroscopic branches.

\subsection{First excited state}

Numerical inspection shows that the main difference between the ground and the first excited states is their behavior under time reversal, i.e., the transformation that changes the sign of all momenta ($k \rightarrow -k$)
\begin{equation}
Ta_kT^\dagger =a_{-k} \, .
\end{equation}
While the ground state is symmetric with regard to time inversion, the first excited state is antisymmetric. For the toy model, the first excited state is very similar to \eqref{GS-toy-model-general} and  \eqref{eq:gs_toy_model} except for the relative sign between $|(N-2m-2l)_{-\pi/2},(2l)_{\pi/2}\rangle$ and $|(2l)_{-\pi/2}, (N-2m-2l)_{\pi/2}\rangle$, which is flipped. As a consequence, the first excited state avoids all configurations for which $N-2m-2l = 2l$. 

\begin{figure}[t!]
	\center
	\includegraphics[width = .45\textwidth]{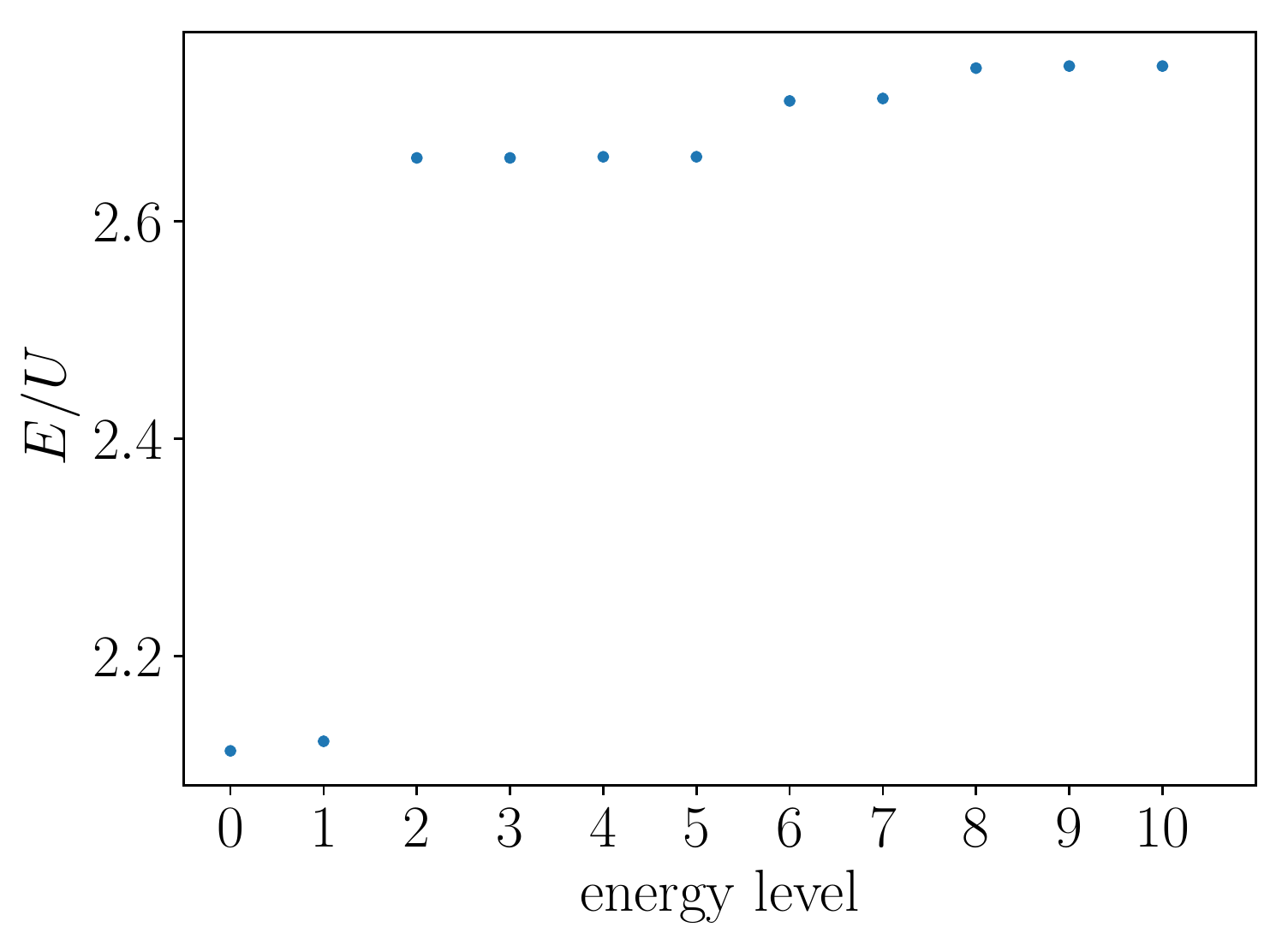}
	\caption{Low-lying excitations of $\Heff$ with $\kappa=0.8$ for 8 particles on a ring of 8 sites. The lowest two energy levels form an almost degenerate doublet, separated by a large energy gap from the rest of the spectrum.}
	\label{Fig:low_exc}
\end{figure}

The energy spectrum for $H_{0.8}$ is shown in Fig. \ref{Fig:low_exc}. The lowest-lying doublet (formed by the similar ground and first excited states) is clearly well-isolated from the higher-lying excited states. 
We also note that in general the values of $|C_{m,l}|$ for a given $m,l$ are different in the ground and the first excited state. This is correlated with the fact that the two lowest-energy states are not degenerate (see Fig. \ref{Fig:low_exc}). Physically, this non-degeneracy is a subtle issue that we discuss in the next section.

\section{Many-body plane waves}
\label{many-body-plane-waves}

Interestingly, we find numerically that the natural orbitals (defined as the eigenstates of the reduced one-particle density matrix) are just plane waves. Of them, the two most occupied ones have momenta $\pm \pi/2$ for both the ground state and the first excited state. Moreover, as already noted, we find numerically that the ground state and the first excited state are, respectively, symmetric and antisymmetric under time reversal. Importantly, this results holds when all coefficients of the ground state wave function in the momentum Fock representation are real. These properties suggest another angle from which to view the relation between the ground state $|\Psi_0 \rangle$ and the first excited state $|\Psi_1 \rangle$, namely, as the symmetric and antisymmetric superposition of two collective plane waves with average momentum $\pm \pi/2$. Specifically, we are led to the following approximate picture for the ground doublet:
\begin{align}
\begin{aligned}
&|\Psi_0\rangle \simeq |C(\pi/2)\rangle \\
&|\Psi_1\rangle \simeq \im |S(\pi/2)\rangle
\end{aligned}
\label{eq:pw_pbc}
\end{align}
where
\begin{align}
	\begin{aligned}
		&|C(k)\rangle \equiv \frac{1}{\sqrt{2}}\left[ |\Phi(k)\rangle + |\Phi(-k)\rangle \right]\\
		&|S(k)\rangle \equiv \frac{-\im}{\sqrt{2}}\left[
		|\Phi(k)\rangle - |\Phi(-k)\rangle
		\right] \, ,
	\end{aligned}
	\label{eq:pw_pbc-generic-k}
\end{align}  
and $|\Phi(k)\rangle =|N_k \rangle$ is a many-body state with all particles in momentum $k$, its wave function being
\begin{equation}
\langle x_1,\ldots,x_N |\Phi(k)\rangle =L^{-N/2} \exp(\im k \sum_i x_i)\, ,
\label{all-in-one-pw}
\end{equation}
where $x_i$ is the space coordinate of the $i-$th boson. Here letters $C$ and $S$ are reminiscent of the $\sin(kx)$ and $\cos(kx)$ wave functions which these states acquire in the single-particle ($N=1$) case. In this limit, $|\Phi(k)\rangle$ becomes the single-particle state of wave function $L^{-1/2}\exp(\im kx)$.

\begin{figure}[t!]
	\center 
	\includegraphics[width=.45\textwidth]{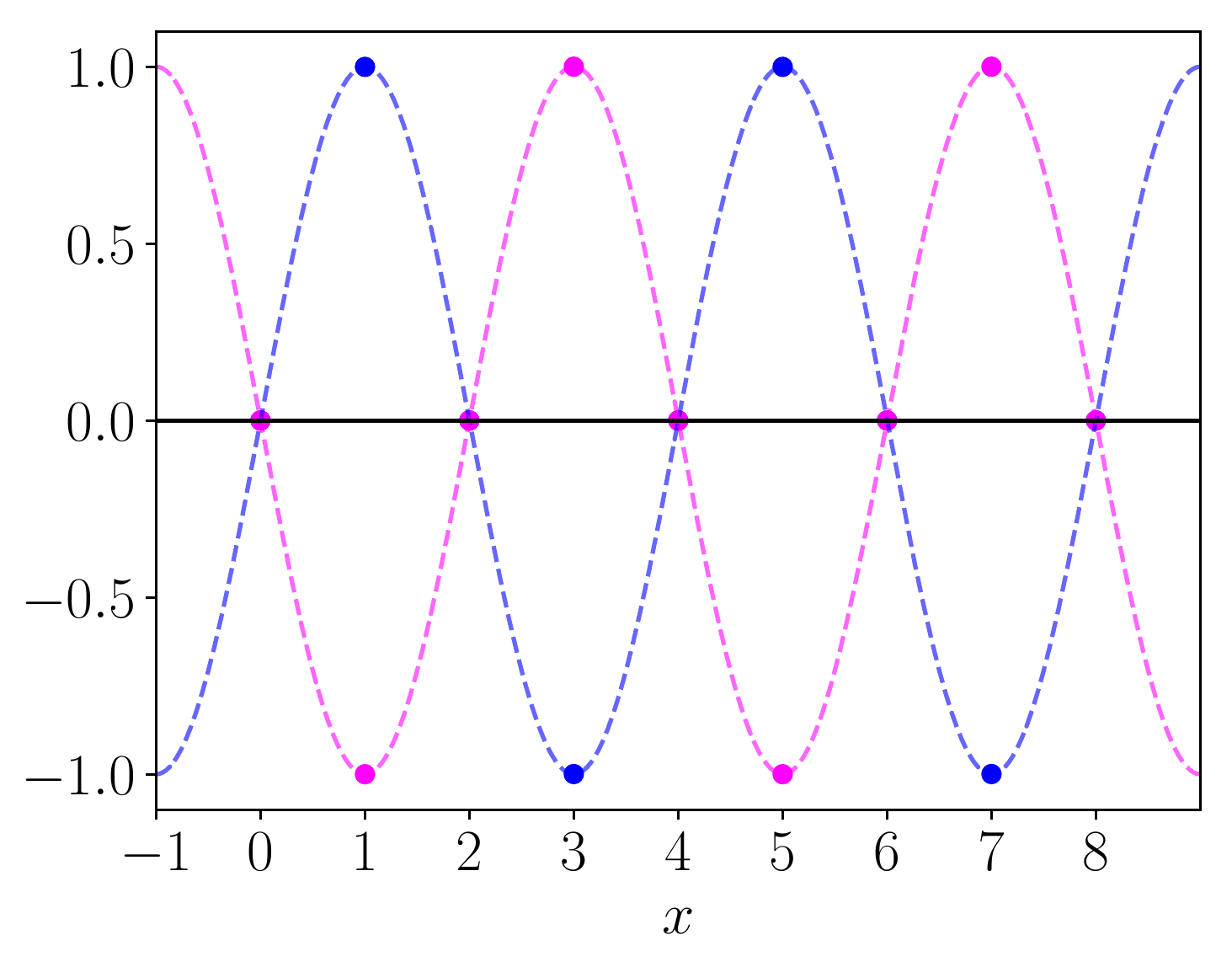}
	\caption{Exact (dots) and ideal (dashed lines) natural orbitals in the ring. The real part of the wave function is plotted.}
	\label{nat_orb_ring}
\end{figure}

In Fig. \ref{nat_orb_ring} we plot the wave function of the numerical and ideal plane wave orbitals for 8 particles on a ring of 8 sites. The difference between the numerical and the ideal results is indistinguishable to the eye.

We can now use the approximate picture \eqref{eq:pw_pbc}-\eqref{eq:pw_pbc-generic-k} as a guide to organize the numerical results. 

Once we have identified the highly correlated nature of the macroscopic occupation of momenta $\pm \pi/2$, as shown in Fig. \ref{Fig:corr_comp}b (which reproduces Fig. 5 of Ref. \cite{pieplow2018generation}), we may wonder whether we can neatly separate the two branches of the cat-like state. If we were dealing with an ideal cat state of the type \eqref{ideal-cat-state} as exemplified in \eqref{eq:pw_pbc-generic-k}, the answer would be easy: the two branches would be $|\Phi(\pm \pi/2) \rangle$, both being states of the form \eqref{GP-state}.

The situation is more complicated when, instead of the ideal cat states \eqref{ideal-cat-state} and \eqref{eq:pw_pbc-generic-k}, we have to deal with numerically obtained states which literally involve thousands of momentum configurations. Considering that all intervening Fock states have total momentum 0 mod $2\pi$, we must rule out the possibility of establishing a criterion to decide to which branch a given momentum configuration contributes. This is even more so if (as will in fact be the case) a given Fock state may contribute to both cat branches.

Equations \eqref{eq:pw_pbc} and \eqref{eq:pw_pbc-generic-k} offer a simple path to identify the two cat branches if $|C(\pi/2)\rangle$ and $|S(\pi/2)\rangle$ are replaced by the true ground and first excited states. One only has to invert \eqref{eq:pw_pbc-generic-k} to propose
\begin{equation}
|\Psi_\pm \rangle= \frac{1}{\sqrt{2}}\left( |\Psi_0 \rangle \pm  |\Psi_1 \rangle  \right)
\label{cat-options-real-case}
\end{equation}
and write
\begin{align}
|\Psi_{0} \rangle &= \frac{1}{\sqrt{2}}\left( |\Psi_+ \rangle +  |\Psi_- \rangle  \right) 
\label{psi-0-options} \\
|\Psi_{1} \rangle &= \frac{1}{\sqrt{2}}\left( |\Psi_+ \rangle -  |\Psi_- \rangle  \right) \,
\label{psi-1-options}
\end{align}
with the orthogonality of $|\Psi_0\rangle$ and $|\Psi_1\rangle$ guaranteeing
\begin{equation}
\langle \Psi_+|\Psi_-\rangle=0 \, .
\end{equation}

\begin{figure}[t!]
	\center 
	\includegraphics[width=.45\textwidth]{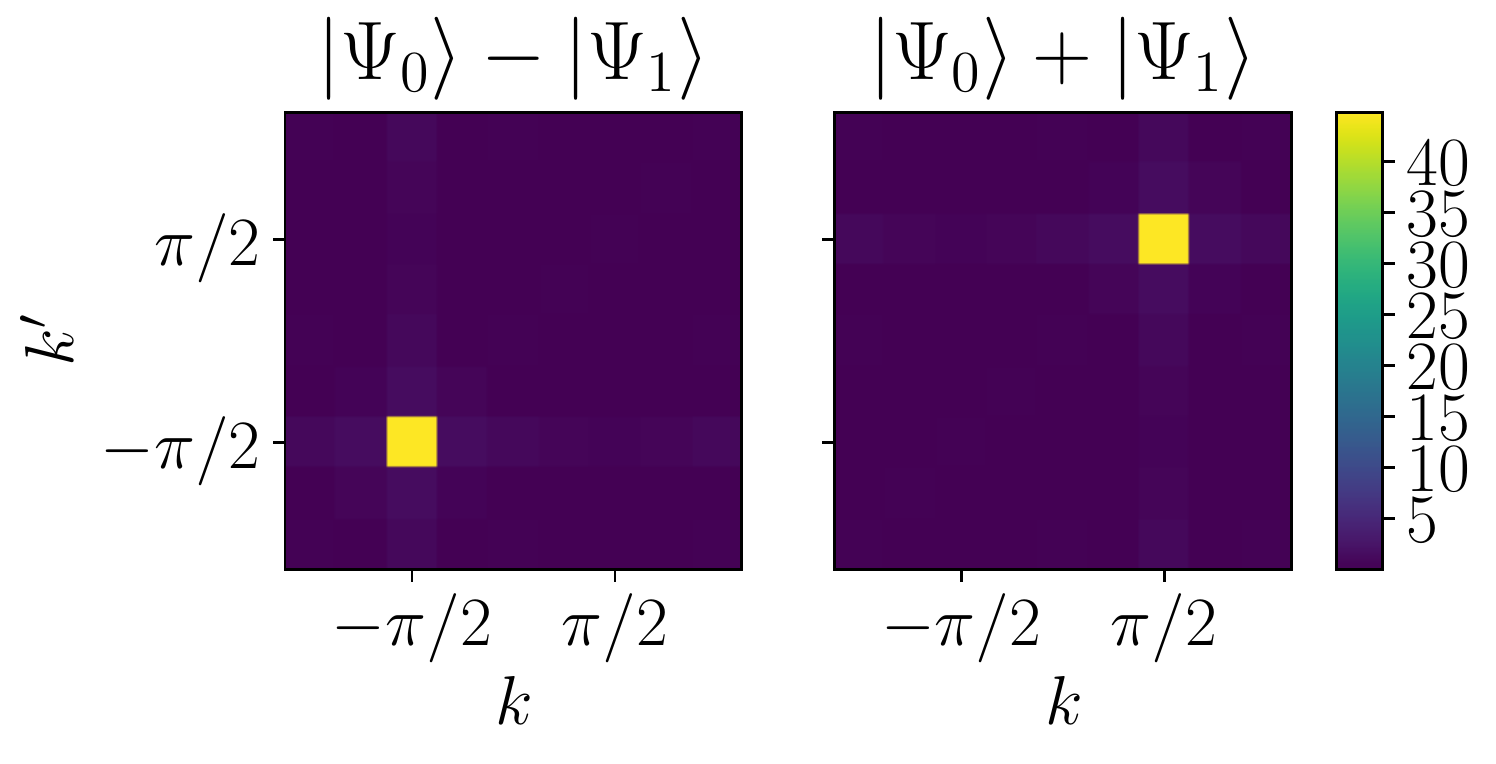}
	\caption{Momentum correlations of the states $|\Psi_\pm \rangle = \frac{1}{\sqrt{2}}(|\Psi_0\rangle \pm |\Psi_1\rangle)$  constructed from the superposition of the ground and first excited states. The isolated peaks clearly suggest that the ground state is formed by two counter-propagating many-body configurations akin to collective plane waves. Here $\kappa=0.7$.}
\label{Fig:pw_gr_1exc-ring}
\end{figure}

In Fig. \ref{Fig:pw_gr_1exc-ring} we show the two-particle momentum density of $|\Psi_\pm \rangle$.
The distinct single peaks at $\pm(\pi/2,\pi/2)$ clearly confirm the adequacy of the criterion \eqref{cat-options-real-case} to cleanly construct the two cat branches. Despite its simple appearance, we emphasize that Fig. \ref{Fig:pw_gr_1exc-ring} shows $\langle n_k n_{k'} \rangle$ for a numerically obtained state $|\Psi_\pm \rangle$
involving thousands of momentum configurations of which $|N_{\pm \pi/2}\rangle$ 
(with all particles in $\pm \pi/2$) is only that with the largest weight. Specifically, for $N=L=8$, we find that the states $|N_{\pm\pi/2}\rangle$ add up to a normalization weight of approximately 50\%.

Remarkably, the decomposition in cat branches such as that shown in \eqref{psi-0-options} with the results of Fig. \ref{Fig:pw_gr_1exc-ring} is also possible for a boson gas between hard walls, as we discuss in the next section. 

Before shifting to the hard-wall case, we finish this section with a note on the nondegeneracy of the ground state doublet shown in Fig. \ref{Fig:low_exc}. In the particular case of $N=1$, \eqref{eq:pw_pbc} yields two degenerate states, since $\cos(\pi x/2)$ and $\sin(\pi x/2)$ are wave functions connected by a symmetry operation, namely, a space translation of one lattice spacing. Interestingly, the same analysis for $N$ particles shows that $|C(\pi/2) \rangle$ and $|S(\pi/2) \rangle$ differ by a global $1/N$ translation. For $N>1$, $1/N$ is less than a lattice spacing and thus such a translation does not yield a degenerate state. However, practical degeneracy is obtained for $N\gg 1$. This argument, developed for noninteracting bosons, provides a semiquantitative explanation of the small but nonzero splitting of the interacting ground state doublet shown in Fig. \ref{Fig:low_exc}. It also suggests that the splitting vanishes in the thermodynamic ($N \rightarrow \infty$) limit.

\subsection{Particle current}
\label{particle-current}

It is tempting to view the branches \eqref{cat-options-real-case} as collective states where many particles ``travel'' with an average momentum of $\pm \pi/2$. However, this picture is invalidated under closer inspection. The unusual character of the effective Hamiltonian \eqref{eq:Heff} yields an also unconventional particle current operator $I_{\kappa}$. If, as a result of a twist in the periodic boundary conditions, all allowed momenta are shifted by an amount $\theta$, the matrix elements in $H_{\kappa}$ change accordingly. The space-averaged particle current operator, $I_{\kappa}= L^{-1}\partial H_{\kappa} /\partial \theta$
can thus be written as
\begin{align}
	\begin{aligned}
	I_{\kappa} = &\frac{U\kappa}{L^2}\sum^{L-1}_{l,m,n,p=0} \mathcal{J}_1 [ 2  \kappa F(k_l,k_m,k_n,k_p) ] \\ 
	&G(k_l,k_m,k_n,k_p)
	a_{k_p}^\dagger a_{k_n}^\dagger  a_{k_m} a_{k_l} 
	\delta_{k_l+k_m,k_n+k_p}
	\, ,
	\end{aligned}
	\label{eq:Ieff}
\end{align}
where ${\cal J}_1$ is the first-order Bessel function and
\begin{equation}
G(k_l,k_m,k_n,k_p)\equiv \sin(k_l)+\sin(k_m)-\sin(k_n)-\sin(k_p) \, .
\label{def-G}
\end{equation}
Clearly, the expectation value of $I_{\kappa}$ vanishes for a state of the type $|\Phi(k)\rangle =|N_k \rangle$ where all particles are in the same momentum $k$. Numerically, we confirm
\begin{equation}
\langle \Psi_{\pm}| I_{\kappa}| \Psi_{\pm} \rangle = \langle \Psi_{0,1}| I_{\kappa}| \Psi_{0,1} \rangle =0 \, .
\label{vanishing-currents}
\end{equation}

However, despite this apparent lack of dynamics, the two branches $| \Psi_{\pm} \rangle$ will behave very differently in a time-of-flight experiment in which the crystal momentum in the lattice becomes the linear momentum in the vacuum as the confining optical lattice is switched off. Moreover, we find that, in the presence of a finite twist in the periodic boundary conditions, the ground state carries a nonzero current (not shown). Interestingly, the nonzero value of the current depends crucially on the presence of the reduction cloud.

\section{Kinetic driving between hard walls}
\label{kinetic-hard-walls}

Here we investigate the effect of kinetically driving a one-dimensional boson system in the presence of hard walls.
With respect to the flat ring scenario studied in Ref. \cite{pieplow2018generation} and in the previous sections, there are similarities but also some key differences. 

The symmetry is greatly reduced and momentum conservation is lost due to lack of translational invariance. The momentum values $0,\pi$ do not play symmetric roles anymore, nor do in general the momenta $k,\pi-k$ with $k\neq \pi/2$, as was the case for the ring. The natural orbitals are not plane waves. The toy model which, derived in the large-$\kappa$ limit, helped us understand much of the physics in a wide range of $\kappa$ values, does not work here anymore.

We no longer can use the plane wave expansion in Eq. \eqref{eq:pw_expansion} to derive the effective Hamiltonian. We must rather introduce stationary waves satisfying the hard-wall boundary conditions.
The usual trick is to extend the lattice by two sites and use stationary waves to expand the creation and annihilation operators in position space [see e.g. Ref. \cite{sols1989theory}]:
\begin{align}
a_{x} = \sqrt{\frac{2}{L+1}} \sum_{l = 1}^{L} \sin (\tilde{k}_l x) a_{\tilde{k}_l}\,\\
a_{\tilde{k}_l} = \sqrt{\frac{2}{L+1}}\sum_{x = 1}^{L} \sin (\tilde{k}_l x) a_x\,
\end{align}
Crucially, the wave vector $\tilde{k}_l$ is defined
\begin{equation}
\tilde{k}_l = \pi l/ (L+1) \, ,
\label{def-k-tilde}
\end{equation}
where $l$ takes $L$ integer values from 1 to $L$, so that $\tilde{k}_l \in (0,\pi)$.
This contrasts with the definition of $k_l$ given in \eqref{define-kp} for plane waves in the ring, with $k_l \in [0,2\pi)$.
Here the momentum density of states is twice as high as for plane waves.

The effective Hamiltonian with hard-wall boundary conditions is therefore
\begin{widetext}
\begin{align}
\begin{aligned}
\Hhw = &\frac{2 U}{(N+1)^2} \sum_{x = 1}^N \sum_{l,m,n,p = 1}^N \sin(\tilde{k}_l x) \sin(\tilde{k}_m x)\sin(\tilde{k}_n x)\sin(\tilde{k}_p x) \times \\
&\hspace{1.5cm}\mathcal{J}_0\big[ 2\kappa
F(\tilde{k}_l,\tilde{k}_m,\tilde{k}_n,\tilde{k}_p) 
\big] a^\dagger_{\tilde{k}_p}a^\dagger_{\tilde{k}_n}a_{\tilde{k}_m}a_{\tilde{k}_l}~,
\end{aligned}
\label{eq:hard_wall_effective}
\end{align}
\end{widetext}
with $F$ defined in \eqref{def-F}.
Its derivation is completely analogous to that of $\Heff$, which was presented in Ref. \cite{pieplow2018generation}. Note that the main difference with respect to \eqref{eq:Heff} is the loss of momentum conservation, and hence the necessary preservation of the 
the residual sum over the position $x$. If the mode functions were plane waves this would yield a simple Kronecker delta, as in \eqref{eq:Heff}. 

The derivation of the toy model for the ring relied on the simple solutions to Eq. \eqref{eq:constraintcond}, which include momentum conservation. An analogous derivation for the hard-wall effective Hamiltonian is too involved to produce a similarly simple Hamiltonian for large $\kappa$. Indeed we find that some properties of the ground state such as its momentum density significantly change for $\kappa > 0.8$ (not shown). By contrast, in the ring case the large-$\kappa$ limit remains consistently smooth for arbitrarily large values of $\kappa$.
	
As to the Luttinger liquid analysis which we made in \cite{pieplow2018generation} for the ring case, we note that the hard-wall boundary conditions drastically alter the superfluid correlations \cite{cazalilla2011one}. This makes it harder to reliably extract Luttinger parameters from small systems.
Since the biggest system we can numerically investigate is 8 particles on 8 sites, we 
will not attempt here to explore the Luttinger liquid properties of the confined boson system.

\subsection{Plane-wave representation}

Although plane waves do not provide a natural basis for quantum particles between hard walls, it is nevertheless possible to introduce them through the transformation:
\begin{align}
a_{x} = \frac{1}{\sqrt{L}} \sum_{l = 1}^{L} e^{\im k_l x} a_{k_l}\, .
\label{truncated-pw-expansion}
\end{align}
In this expansion we implicitly set the plane waves to zero on the fictitious sites $x=0,L+1$ where the hard wall is supposed to be. Because of this we refer to \eqref{truncated-pw-expansion} as a {\it truncated plane wave} expansion. 
This expansion is not useful in deriving the effective Hamiltonian, but will become important later in the physical representation of the results.

The stationary plane waves can be transformed into truncated plane waves via
\begin{equation}
a_{\tilde{k}_l} = \sqrt{\frac{2}{L(L+1)}}\sum_{m=1}^{L} \sum_{x=1}^{L} e^{\im k_{m} x}\sin(\tilde{k}_l x) a_{k_{m}}\, ,
\label{stationary-plane}
\end{equation}
with $\tilde{k}_l$ and $k_m$ are defined as in \eqref{def-k-tilde} and \eqref{define-kp}, respectively.
Unlike in the continuum limit case, the discrete nature of the sum and the fundamental difference between the definitions of $\tilde{k}_l$ and $k_m$ make this transformation non-trivial.
It is also the reason why expressing \eqref{eq:hard_wall_effective} in the plane wave representation does not provide any immediate advantage.

\subsection{Numerical results}
\label{numerical-results}

\begin{figure}[t!]
\center 
\includegraphics[width=.45\textwidth]{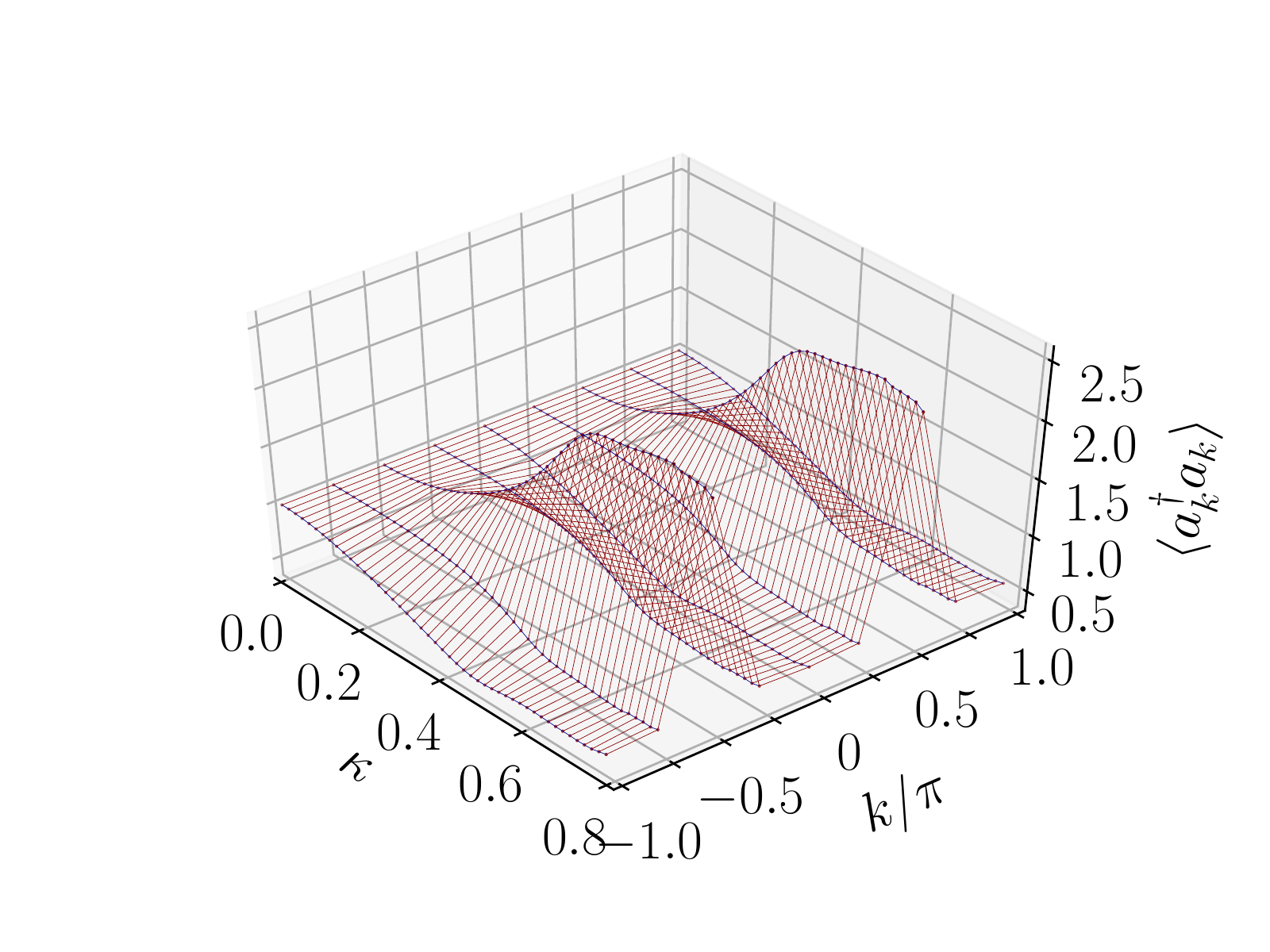}
\includegraphics[width=.45\textwidth]{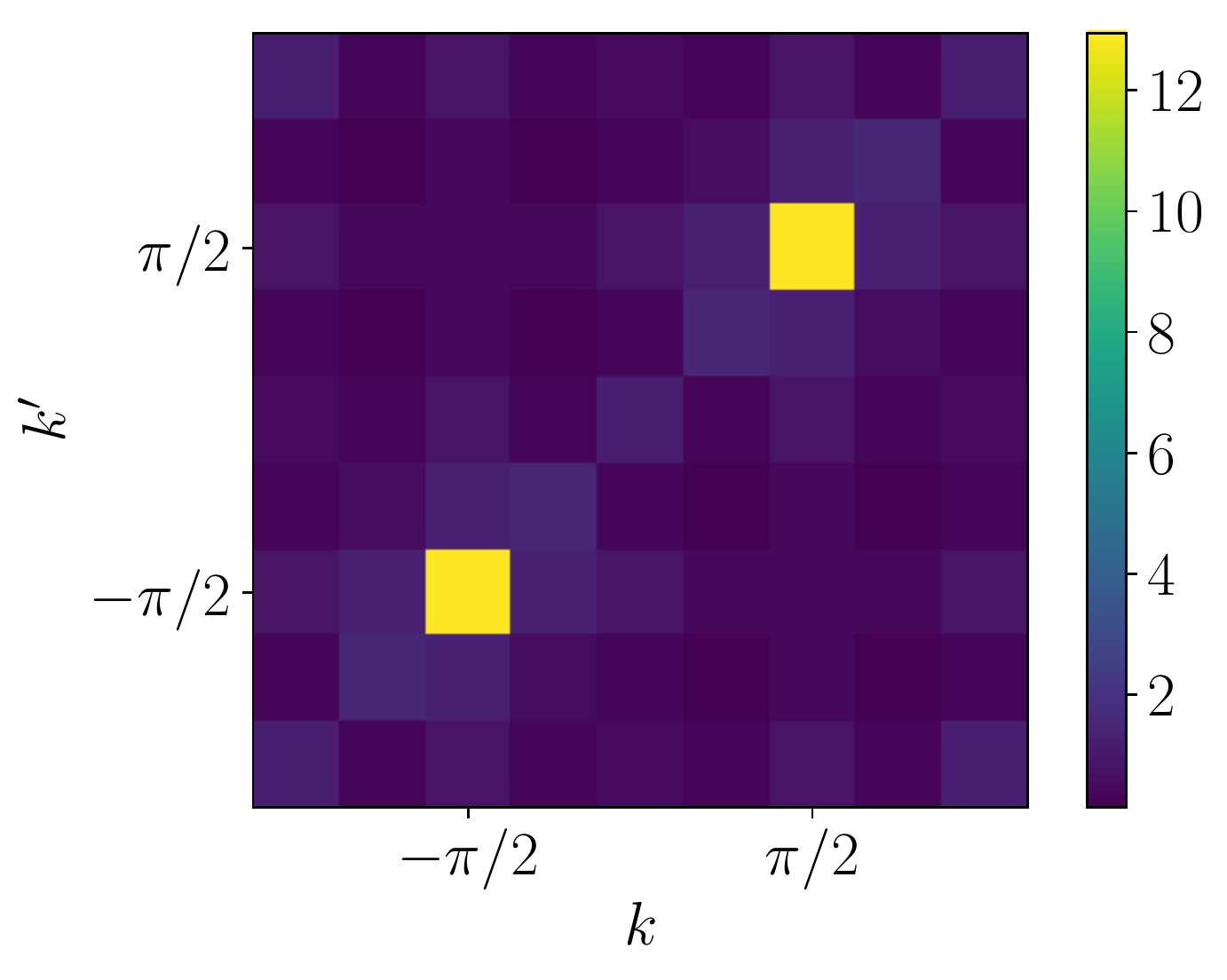}
\caption{Above: Momentum density for the ground state of $\Hhw$ for 8 particles on 8 sites. Just as for the ring the momentum density develops distinct peaks at $k =\pm \pi/2$. For $\kappa=0$ the system is in a Mott state. As $\kappa$ is increased two condensates with non-zero momenta form.
Below: Momentum-momentum correlations at $\kappa = 0.8$ for 8 particles on 8 sites.  The isolated peaks at $\pm (\pi/2, \pi/2)$ indicate that the ground state is cat-like.}
\label{Fig:hw_mom}
\end{figure}

As for the case of periodic boundary conditions, we have studied the momentum density and the momentum-momentum correlations.
To properly compare the ring and hard-wall cases, we work in the 
truncated plane wave representation, which is perfectly reachable through the transformation \eqref{stationary-plane}. Once the ground state has been obtained
in terms of stationary modes, it is easy to investigate the system in the truncated plane-wave representation.  
Interestingly, the properties stay largely intact as compared with the ring case.
As $\kappa$ increases, two distinct peaks form in the momentum density
and the momentum-momentum correlation also shows distinct peaks at $\pm(\pi/2,\pi/2)$ [shown in Fig. \ref{Fig:hw_mom})]. 
Such correlated peaks indicate that the ground state remains Schr\"odinger cat-like even for hard-wall boundary conditions. The ground state is a coherent superposition of two macroscopically distinct states, one with most atoms at positive momenta (centered around $\pi/2$) and the other in the time-reversed configuration.

As in the ring case, we can make use of symmetry to better understand the cat-like structure of the ground and first excited states. Space inversion around the midpoint of the chain amounts to the transformation
\begin{align}
I a_{\tilde{k}_l} I^\dagger = \left\{
\begin{aligned} 
-&a_{\tilde{k}_l} \quad l ~ \text{even}
\\ 
& a_{\tilde{k}_l} \quad l ~ \text{uneven} 
\end{aligned}
\right.
\end{align}
where $I$ is the spatial inversion operator. 
We have checked numerically that the ground state is symmetric under $I$, while the first excited state is antisymmetric. We also see that they have very similar momentum densities and momentum-momentum correlations. 

\begin{figure}[t!]
	\center 
	\includegraphics[width=.45\textwidth]{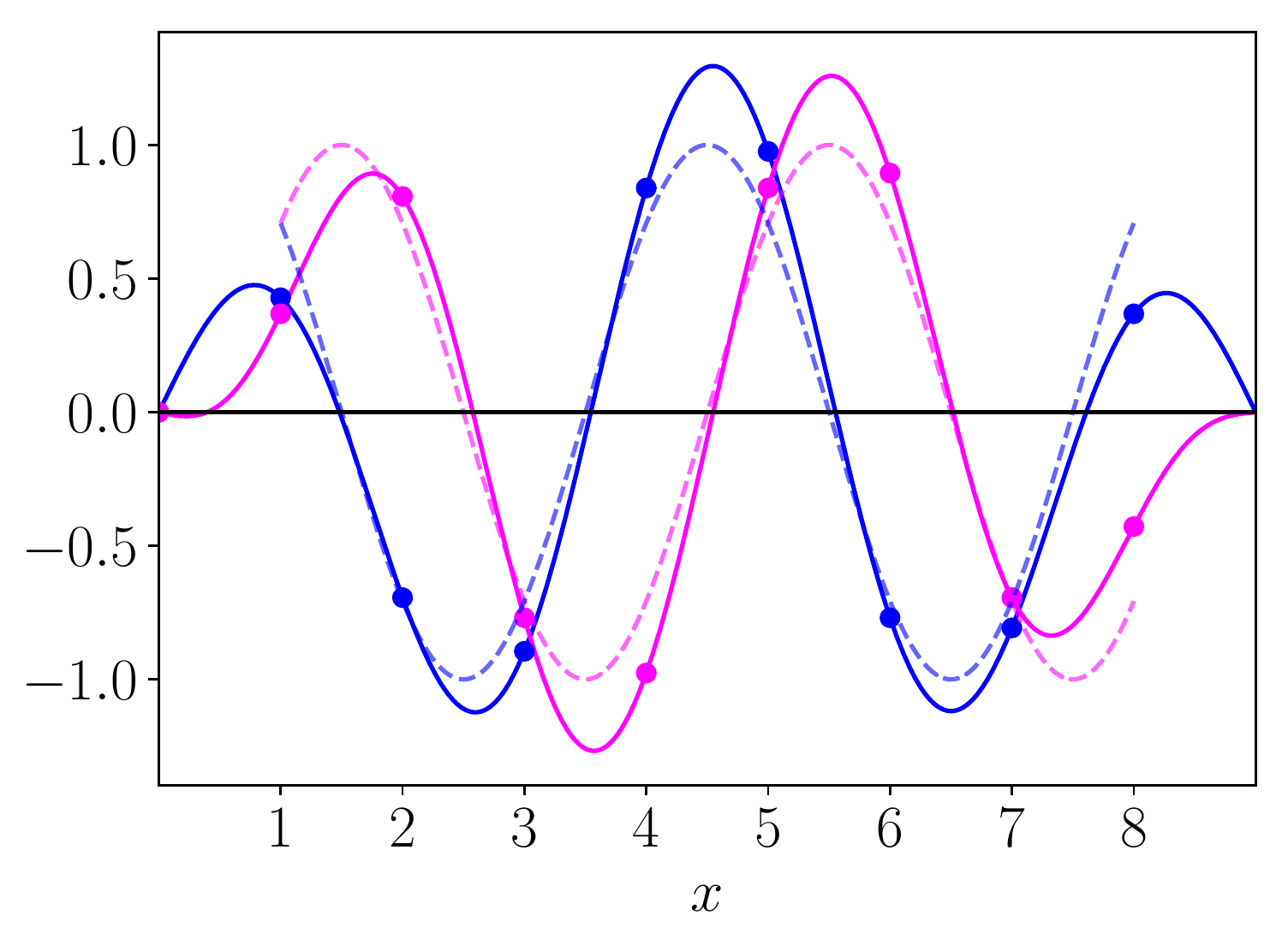}
	\caption{The solid lines show the two most occupied natural orbitals at $\kappa = 0.8$ for 8 particles on 8 sites.
		They are plotted continuously to better compare them with the analytically obtained orbitals from the ansatz in \eqref{eq:schr_sketch_hw_gs}. The points indicate the value of the wave function at the real sites. They show very good agreement in the center of the lattice. At the edges they start to disagree since the truncated waves are highly discontinuous at the edges while the stationary waves are not.}
\label{Fig:nat_orbs}
\end{figure}

As we did for the ring, we can translate this into an approximate picture for the wave function of the ground and first excited states. Specifically, we propose
\begin{align}
|\Psi_0\rangle &\simeq  |C(\pi/2)\rangle 
\label{eq:schr_sketch_hw_gs}
\\
|\Psi_1\rangle &\simeq  |S(\pi/2)\rangle \, ,
\label{eq:schr_sketch_hw_ex}
\end{align}
with the coordinates in \eqref{all-in-one-pw} referred to the midpoint between the walls.
The adequacy of this approximation for some purposes can be inferred from an analysis of the approximate and numerically-exact natural orbitals. It is possible to work out the reduced one-particle density matrix for $|C(k)\rangle$ and $|S(k)\rangle$. One obtains
\begin{align}
\rho^{(1)}(x,x') &= \frac{1}{L} \cos[k(x - x')]\\
&= \frac{1}{L}[\cos(k x)\cos(k x')+\sin(k x)\sin(k x')]\, ,
\end{align}
from which we conclude that the natural orbitals are
\begin{equation}
\phi_0(x) = \cos(k x)\, ,\quad \phi_1(x) = \sin(k x)\, .
\label{approx-nat-orb}
\end{equation}
We can compare them with the natural orbitals obtained numerically. Setting $k = \pi/2$, we find that they are indeed very similar. Figure \ref{Fig:nat_orbs} shows a direct comparison between the degenerate most occupied orbitals for 8 particles on 8 sites and the approximate natural orbitals in \eqref{approx-nat-orb}.
We notice that the numerically exact state obeys the boundary conditions, so its orbitals go more gradually to zero close to the boundaries. One could ameliorate this by multiplying the truncated plane waves implicit in \eqref{eq:schr_sketch_hw_gs} and 
\eqref{eq:schr_sketch_hw_ex} by an envelope function that smoothly goes to zero at the walls. This is also what in practice breaks the symmetry between $0$ and $\pi$, since the envelope function's discontinuity at the walls will imbalance the momentum distribution. 

The foregoing analysis of symmetries and natural orbitals strongly supports the relevance of the approximate picture \eqref{eq:schr_sketch_hw_gs} proposed above. As we did for the ring, we can invert the terms and use the relations \eqref{eq:pw_pbc-generic-k} to identify the cat branches between hard walls.

The fact that the ground state and first excited states are, respectively, even and odd under space inversion (always within the convention that, when expanded into momentum Fock states, the ground and first excited state have only real coefficients) strengthens the argument that our ground and first excited states are similar to \eqref{eq:schr_sketch_hw_gs} and \eqref{eq:schr_sketch_hw_ex}. 

\begin{figure}[t!]
\center 
\includegraphics[width=.45\textwidth]{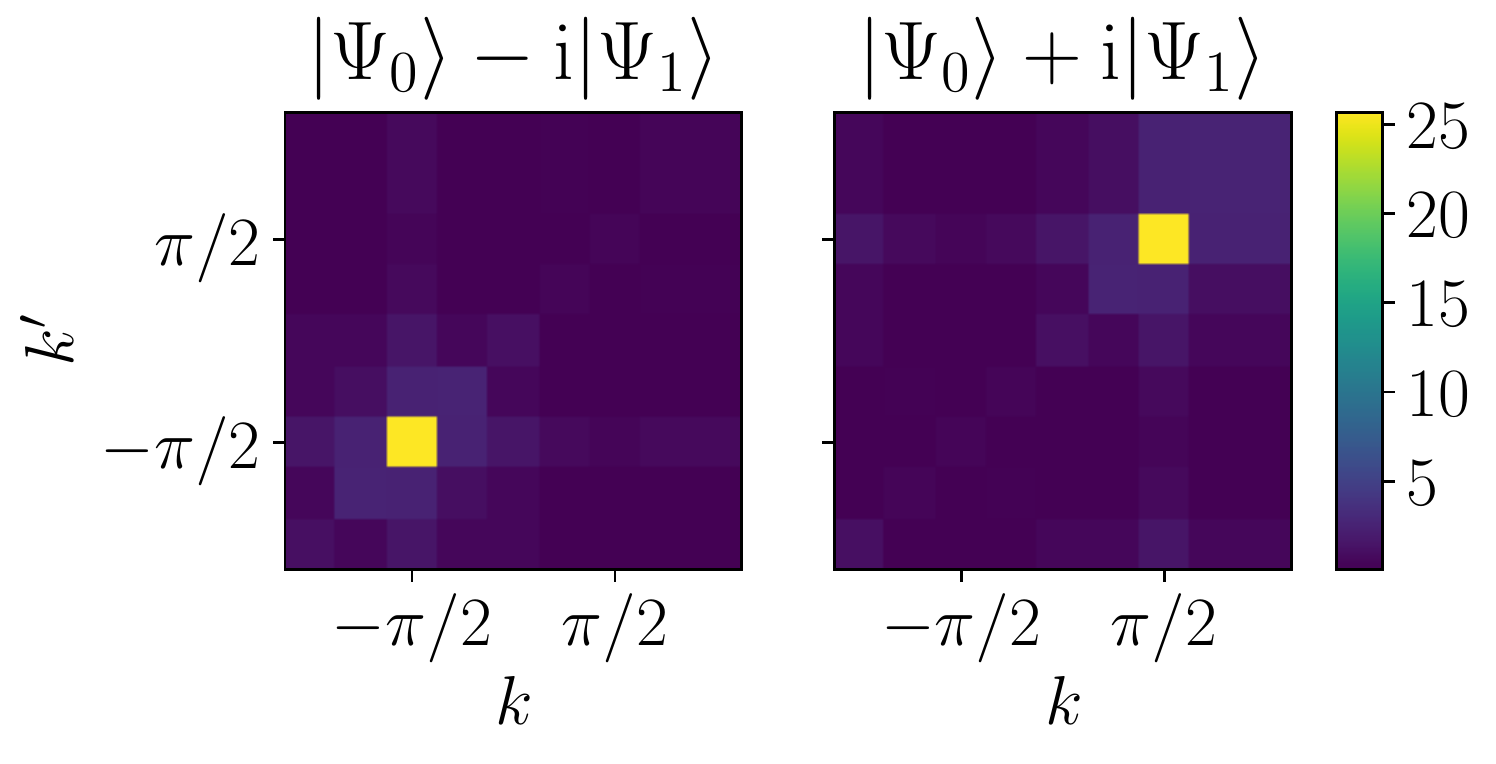}
\caption{Momentum correlations of the states $|\Psi_\pm \rangle$  constructed from the superposition of the ground and first excited states [see \eqref{options-hw-case}] for $\kappa=0.8$. The isolated peaks clearly suggest that one can view the ground state as the superposition of two counter-propagating collective plane waves between the walls.}
\label{Fig:pw_gr_1exc}
\end{figure}

If we understand that the ground state and the first excited state are, respectively, cosine-like and sine-like (with respect to the midpoint) we can expect
the cat branches to be:
\begin{equation}
|\Psi_{\pm}\rangle = \frac{1}{\sqrt{2}}(|\Psi_0\rangle \pm \im |\Psi_1\rangle) \, .
\label{options-hw-case}
\end{equation} 
This expectation is confirmed in Fig. \ref{Fig:pw_gr_1exc}, where the numerical momentum-momentum correlation is shown for the branches $|\Psi_{\pm}\rangle$. As argued for the ring, the orthogonality of $|\Psi_{0}\rangle$ and $|\Psi_{1}\rangle$ guarantees $\langle \Psi_+|\Psi_{-}\rangle=0$.

The outcome is that for the hard-wall case the ground and first excited states can again be written as
\begin{align}
|\Psi_{0}\rangle &= \frac{1}{\sqrt{2}}(|\Psi_+\rangle + |\Psi_-\rangle)\\
|\Psi_{1}\rangle &= \frac{-\im}{ \sqrt{2}}(|\Psi_+\rangle - |\Psi_-\rangle) \, ,
\end{align}
where $|\Psi_\pm \rangle$ represent complex but strictly orthonormal many-body configurations with a preferential occupation of momenta $\pm \pi/2$. We identify them with the two orthonormal branches of the cat state.
As for the ring case, we remark that the states $|\Psi_{\pm}\rangle$ yielding the simple-looking result of Fig. \ref{Fig:pw_gr_1exc} actually involve thousands of momentum configurations.

Like for the ring case (see Section \ref{particle-current}), one might naively interpret the branches \eqref{options-hw-case} as traveling many-body states. We also find here that the expectation value of the current operator [see Eq. \eqref{eq:Ieff}] 
vanishes in the ground doublet, with expressions identical to \eqref{vanishing-currents}.

\begin{figure}[t!]
	\center 
	\includegraphics[width=.45\textwidth]{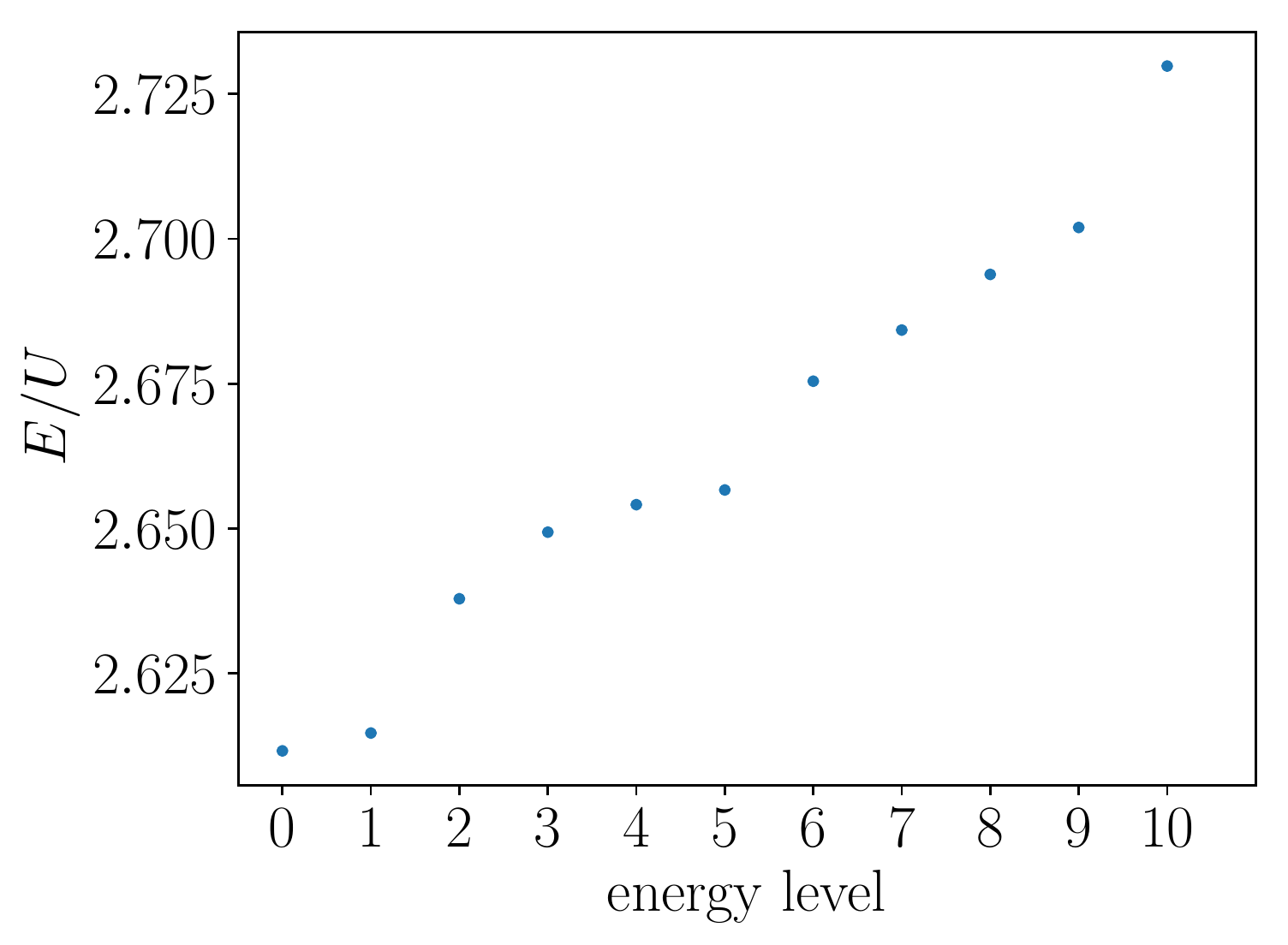}
	\caption{Energy levels of ground and lowest-lying excited states for $\kappa=0.8$ with hard-wall confinement. The gap between the lowest two states and the rest of the spectrum is considerably smaller than for the ring-case (Fig. \ref{Fig:low_exc}).}
\label{Fig:low_en_exc_hw}
\end{figure}

For completeness, we show in Fig. \ref{Fig:low_en_exc_hw} the spectrum for the hard wall system. The lowest doublet is separated from the higher-lying excitations by a smaller energy interval than in the ring case. The comparison of the various energy differences in the ring and in the hard-wall cases takes us to the question of the fragility of the cat-like correlations against finite-time state preparation, a problem which we address in the section~\ref{state-preparation}.

\subsection{Harmonic confinement}

We have so far considered the case of a box potential, that is,
a potential that is zero within two hard-wall boundaries. Although this type
of trapping has been used in experiment \cite{gaunt2013bose}, it is much
more common to use a parabolic trap. To see how the presence of
such a potential may modify the results we obtain, we add a potential
term to the driven Bose-Hubbard model
\begin{equation} 
V = V_0 \sum_j \left( x_j - x_0 \right)^2 n_j \ ,
\end{equation}
where $x_0$ is the centre of the box and $n_j$ is the standard
number operator. As before, we prepare the system in a perfect
Mott state, and then add the above potential while
slowly increasing the amplitude of the time-dependent driving.

For a weak trapping potential, $V_0=0.01 U$ and $J=U$, the momentum density function
and the occupation of the natural orbitals strongly resemble
the case of the flat trap ($V_0 = 0$). As $\kappa$ is increased
from zero, two natural orbitals become macroscopically occupied,
while two peaks centered on momenta $\pm \pi / 2$ appear in the
momentum density function. This provides strong evidence for
the formation of a similar Schr\"odinger cat state. The appearance of the two peaks remains
true for higher values of the trap curvature ($V_0/U = 0.04$ and $0.08$)
This is also true for random potentials; the cat features survive
as long as the amplitude of the disorder potential is not large
enough to localize the particles (not shown).

\section{Measures of ``cattiness''}
\label{measures-of-cattiness}

The question remains of how to quantify the quality of a cat-like state, which may loosely be referred to as ``cattiness'' \cite{Everitt2014}.
A number of cat quality measures have been proposed \cite{leggett2002testing,Jeong2015,frowis2015linking}. 
Quantitative measures tend to reward both the purity (branch orthogonality) and the size of the system.
From the summary of Ref.~\cite{Jeong2015} we may distinguish 
several approaches chosen to characterize the quality of an MQS state: Definition of the effective size of the superposition by comparing it to a cat state made from two orthogonal modes \cite{dur2002effective,Marquardt2008,lee2011,frowis2012measures}, 
size of the relative particle-number fluctuations \cite{hoyip2000,ho2004}, entanglement measures taken from quantum information theory \cite{leggett1980macroscopic,carr2010,mazzarella2011}, magnitude of the off-diagonal correlations to distinguish mixed states from cat states \cite{cavalcanti2006signatures,cavalcanti2008,haigh2010demonstrating,opanchuk2016quantifying}, and more measurement-based approaches \cite{Bjrk2004,korsbakken2007measurement,korsbakken2009electronic,frowis2013certifiability,sekatski2014size}.    

In this section, in order to characterize the ground states found in our calculation, we adopt the pragmatic, measurement-based definition of cat quality introduced in \cite{korsbakken2007measurement}. Since we deal with pure many-body states involving many one-particle modes, this measure is
particularly helpful to us because its general character is not restricted to the case of two modes.
The measure in \cite{korsbakken2007measurement} is convenient for us also because we know the branches precisely.
It attempts to quantify how well the two branches $A$ and $B$ of a superposition
\begin{equation}
|\Psi \rangle = |A\rangle + |B\rangle
\label{A-plus-B}
\end{equation}
can be distinguished after a measurement. 
More specifically it gauges the maximal probability of successfully inferring
the two $n$-particle reduced density matrices (n-PRDMs) $\rho_A^{(n)}$ and $\rho_B^{(n)}$ \cite{fuchs1999cryptographic} related to the respective branches after an $n$-particle measurement:
\begin{equation}
P^{(n)} = \frac{1}{2} + \frac{1}{4}||\rho^{(n)}_A - \rho^{(n)}_B||~.
\label{eq:KCmeasure}
\end{equation}
where $||\cdot||$ is the trace norm $||\rho|| = \sum_i |\lambda_i|$ with $\lambda_i$ the eigenvalues of $\rho$. 
If both density matrices are identical, then the probability of correctly inferring the branch from an $n$-particle measurement is $P = 1/2$. This means one has not learned anything about either branch. For a perfect cat state $P^{(1)} = 1$, which means an immediate ``collapse" takes place after the first particle measurement.
The more particles it takes to differentiate the branches, the smaller is the effective size of the superposition. 

Here we simply evaluate \eqref{eq:KCmeasure} without linking it to an effective cat-size through a value of confidence as the authors of Ref.~\cite{korsbakken2007measurement} did. We only consider up to two particle measurements.  
In the plane-wave representation, both branches are sharply defined by the splitting in Eq. \eqref{psi-0-options}.

\begin{figure}[t!]
	\center
	\includegraphics[width=.9\columnwidth]{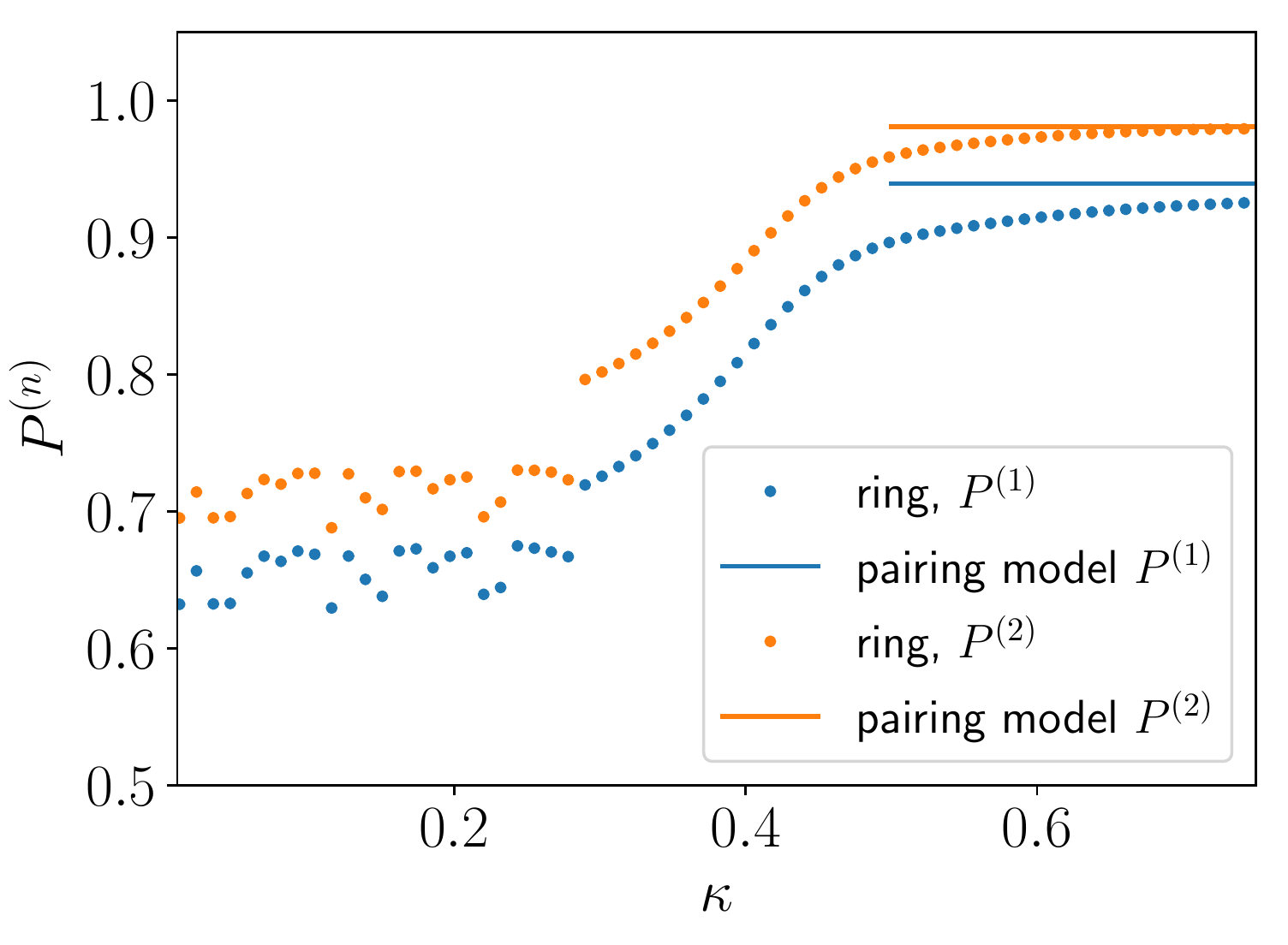}
	\includegraphics[width=.9\columnwidth]{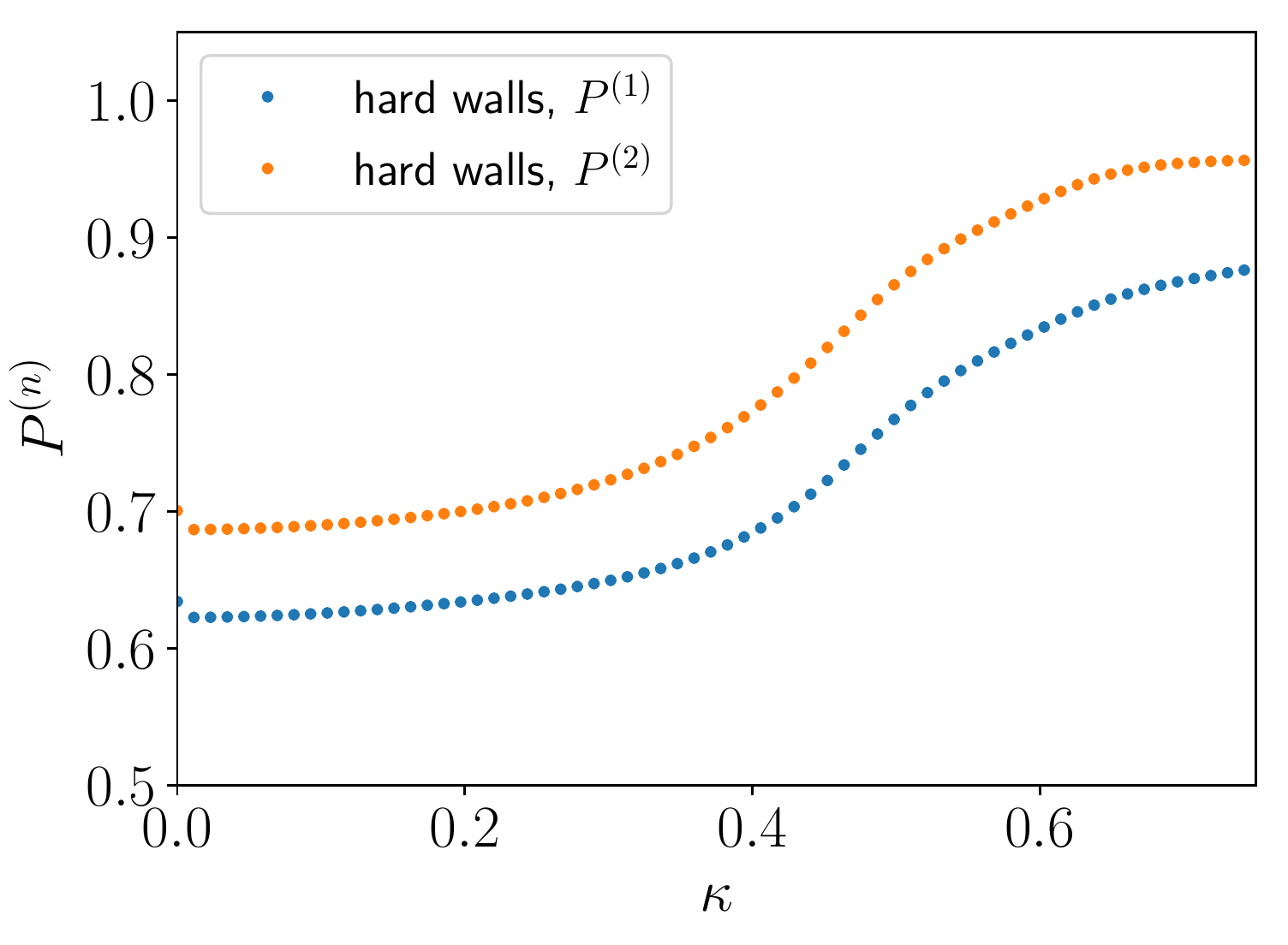}
	\caption{Figure of merit based on the probability $P^{(n)}$ of knowing if the state collapsed into either branch after an $n$ particle measurement. Top: ring, bottom: hard walls. The apparent discontinuity in the ring at $\kappa \simeq 0.3$ is due to a discontinuity in the behaviour of the first excited state. The probabilities to infer the cat state for the hard-wall case are consistently lower than for the ring.}
	\label{Fig:cirac_cat}
\end{figure}

The properly normalized components of the one-PRDM and two-PRDM are
\begin{align}
&\rho^{(1)}_{il}  = \frac{1}{N}\langle a_{k_i}^\dagger a_{k_l} \rangle
\\
&\rho^{(2)}_{ijlm} = \frac{1}{N(N-1)}\langle a_{k_i}^\dagger a_{k_j}^\dagger a_{k_l} a_{k_m} \rangle
\end{align}
In Fig. \ref{Fig:cirac_cat} we present $P^{(n)}$ as a function of $\kappa$ for $n=1,2$. 
As $\kappa$ grows, it is evident that the probability of inferring $A$ or $B$ rises in both the ring and the hard walls. For $\kappa< 0.75$ the biggest probabilities reached for the ring are $P^{(1)} = 0.93$ and $P^{(2)} = 0.98$, whereas for the hard walls  $P^{(1)} = 0.88$ and $P^{(2)} = 0.95$.
In the ring one can see that the probabilities closely approach their theoretical limit,
$P^{(1)} = 0.94$, $P^{(2)} = 0.98$, provided by the pairing model. Even in the large-$\kappa$ limit, the ground state is not a perfect cat, since the pairs in the reduction cloud are shared by both condensate branches. This effect seems to survive in the thermodynamic limit, as we argue below.

As expected, the probability for collapsing the state after a two-particle measurement is consistently higher in both the ring and the hard-wall case. 
The apparent discontinuity in the ring for low $\kappa$ stems from a discontinuity of the first excited state, which does not develop its peaks at $\pm \pi/2$ in the momentum density gradually (not shown).
\begin{figure}[htb!]
	\center
	\includegraphics[width=.9\columnwidth]{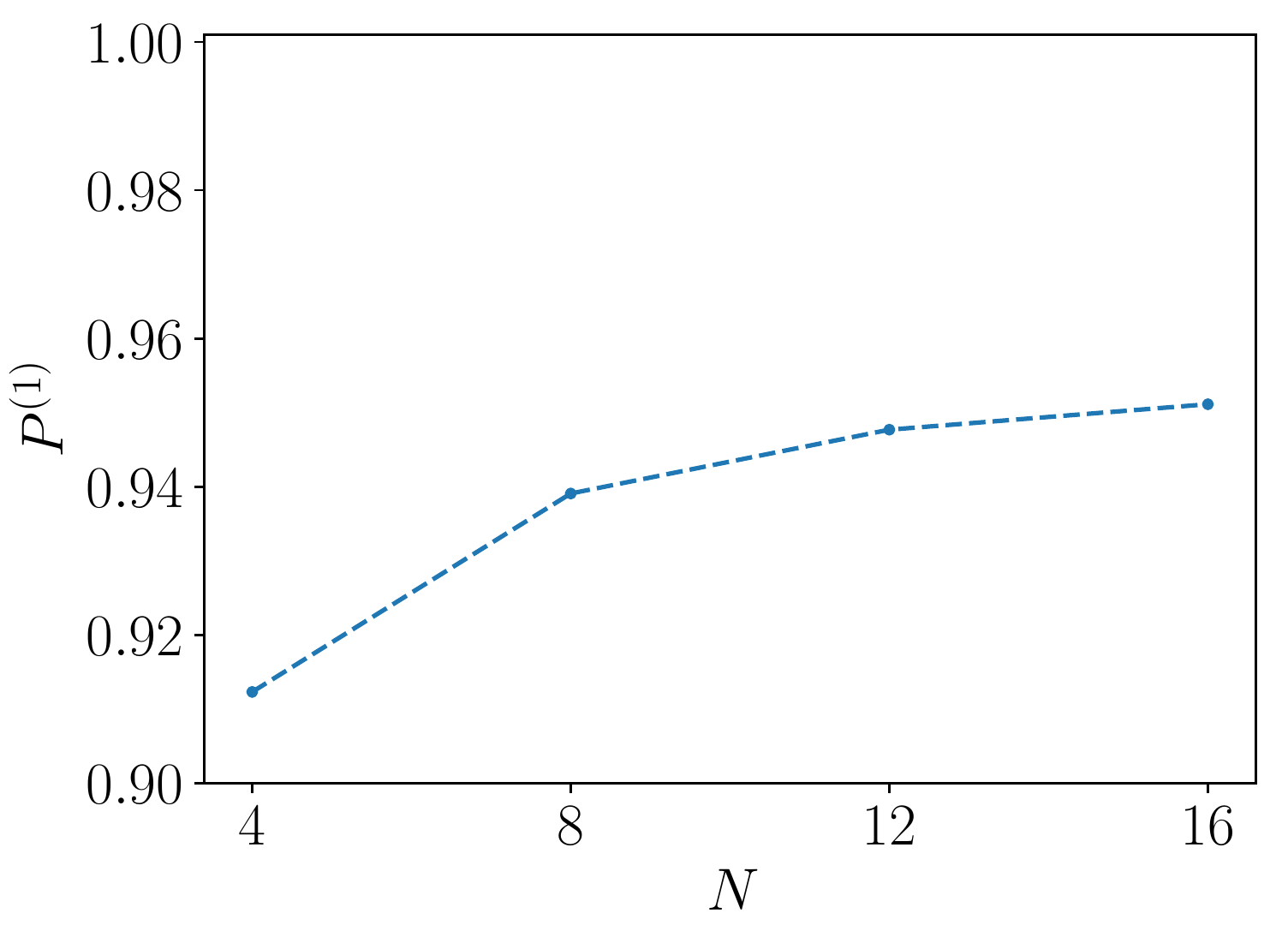}
	\caption{Scaling of $P^{(1)}$ with system size for the pairing model. For up to $16$ particles on $16$ sites the probability steadily increases.}
	\label{Fig:cirac_Toy_Scaling}
\end{figure}

Finally in Fig. \ref{Fig:cirac_Toy_Scaling} we look for the scaling of $P^{(1)}$ with growing system size. The pairing model predicts that $P^{(1)}$ increases with $N$ but saturates to a value below unity for $N\rightarrow \infty$ due to the presence of the reduction cloud. 
This result suggests that the weight of the reduction cloud saturates to a finite value in the thermodynamic limit.

Now we define a complementary figure of merit $C$ to quantify cattiness:
\begin{equation}
C \equiv \sum_k p_k (|\langle A|\xi_k\rangle|-|\langle B|\xi_k\rangle|)^2
\label{measure-C}
\end{equation}
where $|A \rangle$ and $| B \rangle$ are normalized, and
\begin{equation}
p_k = \langle \Psi| n_k | \Psi\rangle/N~
\label{eq:mom_prob}
\end{equation}
is the probability of finding momentum $k$ in a one-particle measurement, and
\begin{equation}
|\xi_k\rangle = \frac{n_k}{\mu_k^2}|\Psi\rangle~,
\label{eq:xik}
\end{equation}
with 
$\mu_k^2 \equiv \langle\Psi|n_k^2 | \Psi \rangle$, is the normalized many-boson state after projecting out configurations with momentum $k$ unoccupied.
The cattiness $C$ is an intuitive measure tailored to our system. As in Ref. \cite{korsbakken2007measurement}, we have looked for a figure of merit based on a measurement procedure. The idea is to infer the quality of the superposition by correlating it with the extent to which $|\xi_k\rangle$ overlaps with $|A\rangle$ or $|B\rangle$, i.e., with our ability to predict the outcome of the second one-particle measurement from the result of the first measurement.

For an ideal cat, $C = 1$. For a non-cat state ($|A\rangle=|B\rangle$ in \eqref{A-plus-B}), $C=0$.
\begin{figure}[t!]
	\center
	\includegraphics[width=.9\columnwidth]{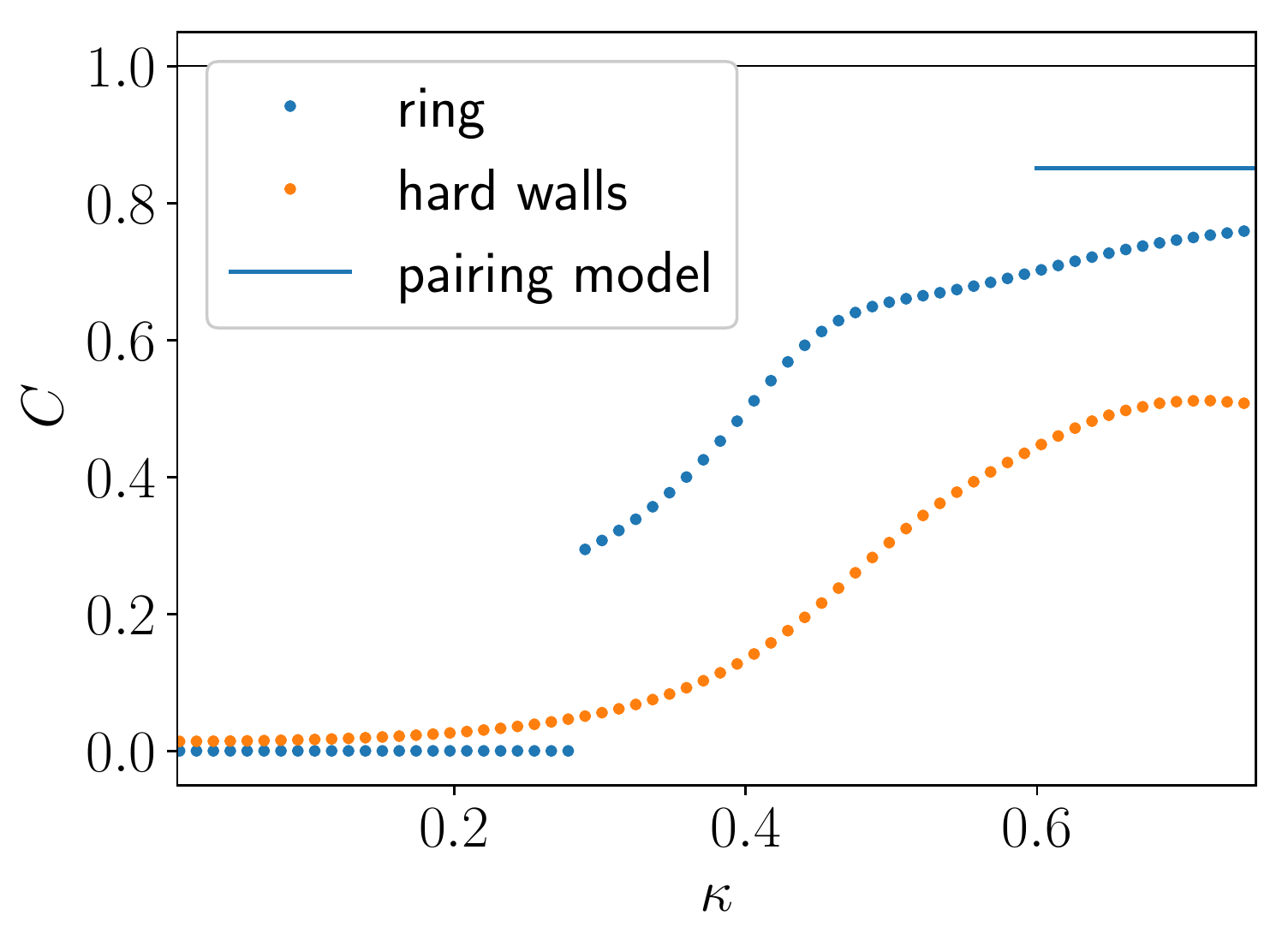}
	\caption{Figure of merit for ``cattiness'', $C$, defined in Eq. \eqref{measure-C}. As before, the discontinuity in the ring result arises from the behaviour of the first excited state. $C<1$ for large $\kappa$ because both branches share the reduction cloud. The largest values reached are: $C = 0.79$ for the ring, $C = 0.85$ for the pairing model and $C = 0.51$ for hard walls.}
	\label{Fig:ad_hoc_merit}
\end{figure}
In Fig. \ref{Fig:ad_hoc_merit} we show the results for the ring and hard-wall boundary conditions. Again, the discontinuity for the ring stems from a discontinuity in the nature of the first excited state. In the Mott insulating regime, $C$ is very close to zero and increases notably after the transition. As for \eqref{eq:KCmeasure} we associate the fact that $C<1$ for large $\kappa$ to the presence of the reduction cloud, i.e., the multi-mode nature of the ground state. 

The quantitative measure \eqref{measure-C} is particularly simple to apply when only two modes are involved. As an illustration, in 
Appendix \ref{app:merit} we explicitly calculate $C$ for a state of the form
\begin{align}
|\Psi\rangle = \frac{1}{K}\left(|A\rangle + |B\rangle \right)
\end{align}
where $K$ normalizes $|\Psi \rangle$ and
the branches are constructed from non-orthogonal single particle orbitals:
\begin{align}
&|A\rangle = \frac{1}{\sqrt{N!}}(a^\dagger)^N |{\rm vac}\rangle~,\\
&|B\rangle = \frac{1}{\sqrt{N!}}[\cos(\theta) a^\dagger + \sin(\theta) b^\dagger  ]^N|{\rm vac} \rangle ~.
\end{align}
For $\theta \in (0,\pi/2)$, we find that $C$ nicely interpolates between 0 and 1.

\section{State preparation}
\label{state-preparation}

\begin{figure}
\center
\includegraphics[width = .49\textwidth,clip=true]{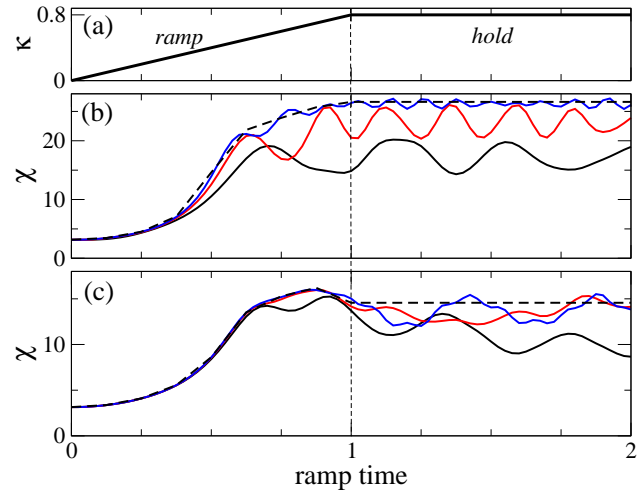}
\caption{The full time-dependent system is initialized in a Mott state, and $\kappa$ ramped linearly from 0 to a value of 0.8 over a time-interval (the ramp-time), and then held constant
for the same time interval
(a) Time-dependence of the driving parameter, $\kappa$.
(b) Cat correlation $\chi$ for a ring system.
The black line is for a ramp-time of $1600 T$, the red line for $3200 T$ and the
blue line for $6400 T$. As the ramp-time is increased, the final value of $\chi$ becomes closer
to that of the system's true ground state (dashed curve), and the oscillations reduce
in amplitude, indicating that the cat state is being prepared with greater fidelity.
(c) As in (b) but for a hard-wall system. The final value of $\chi$ is lower, and
even for the slowest ramp-time, the oscillations in $\chi$ remain significant. Much
slower ramps would be needed to prepare the cat state with adequate fidelity.
Physical parameters: $U=1$, $\omega = 50$.
}
\label{Fig:cat_contrast_adiabatic_preparation}
\end{figure}

We now consider how the cat state can actually be prepared in an experiment. As it
is the ground state of the system, we might expect to naturally fall into this state
as the system is cooled. However, the cat state consists of a superposition of two
branches which are degenerate in energy and yield a splitting that 
vanishes in the thermodynamic limit. As a consequence, cooling
the system towards zero temperature will just  result in a classical mixture of the two branches
\cite{daki2017}. 
We will instead look at the feasibility of preparing the cat state by adiabatic manipulation, 
as considered in \cite{moore2006,zoller1998}.
Working with the full time-dependent Hamiltonian \eqref{eq:driven_BH}, we initialise the
system in the Mott state, with one particle occupying each lattice site, and slowly ramp
$\kappa$ up over many thousands of driving periods from an initial value of zero to 
a final value of $\kappa = 0.8$, as shown in Fig. \ref{Fig:cat_contrast_adiabatic_preparation}a. When this final value is reached, we then hold $\kappa$
constant.

To give an indication of the degree to which the instantaneous state of the system 
exhibits cat correlations during this process, we evaluate the two-particle reduced density matrix
at momenta $\left( \pi/2, \pi/2 \right)$,
that is, we compute $\chi \equiv \langle n_{\pi/2} n_{\pi /2 }\rangle$.
As we have seen earlier, this quantity has a large value when the system is in a cat-like state and is small otherwise. So it acts
as a reasonable alternative figure of merit to characterize the 
expected cat nature of the
state we reach upon slowly ramping up the driving amplitude.

\begin{figure}
\center
\includegraphics[width = .45\textwidth]{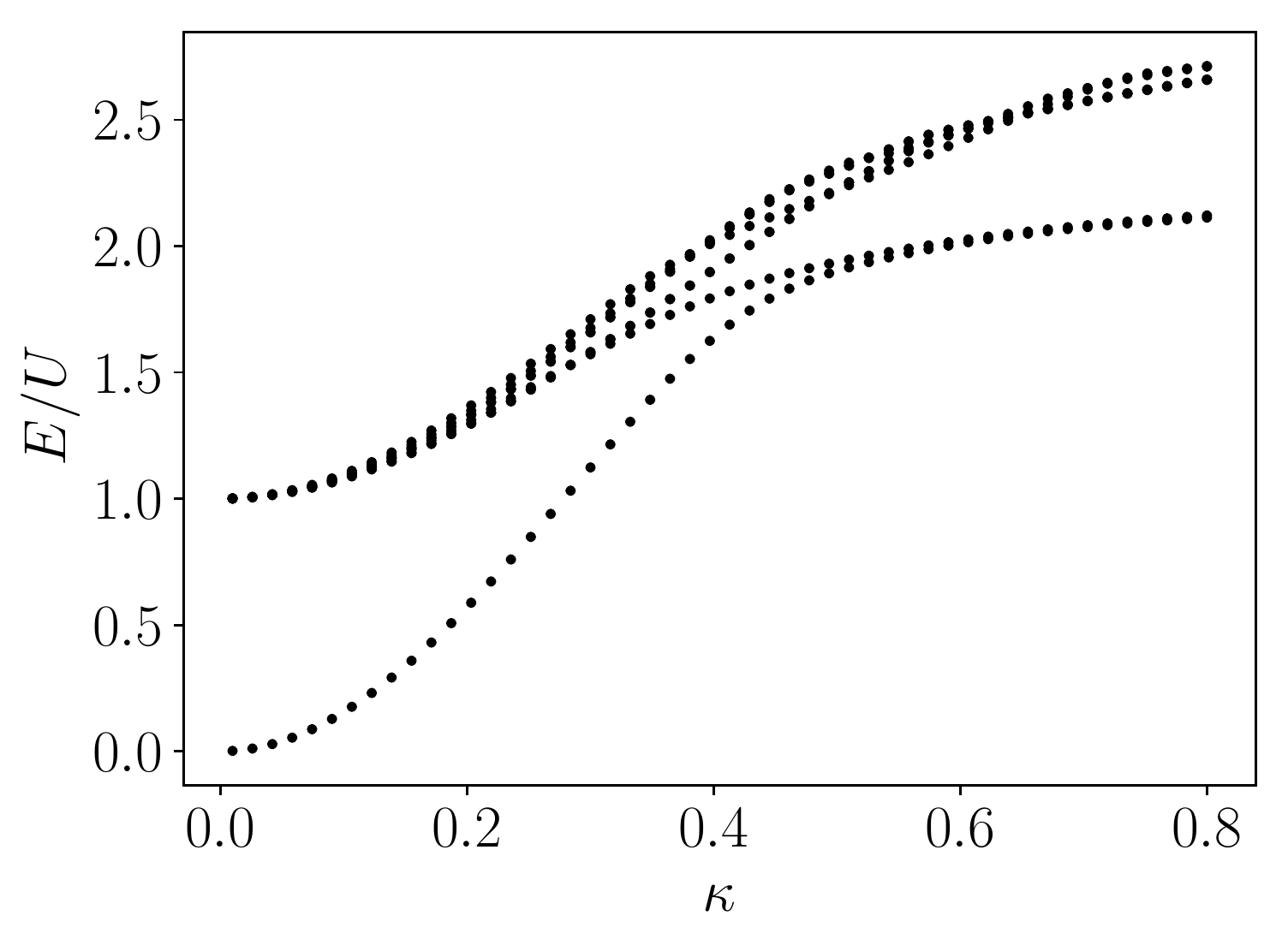}
\includegraphics[width = .45\textwidth]{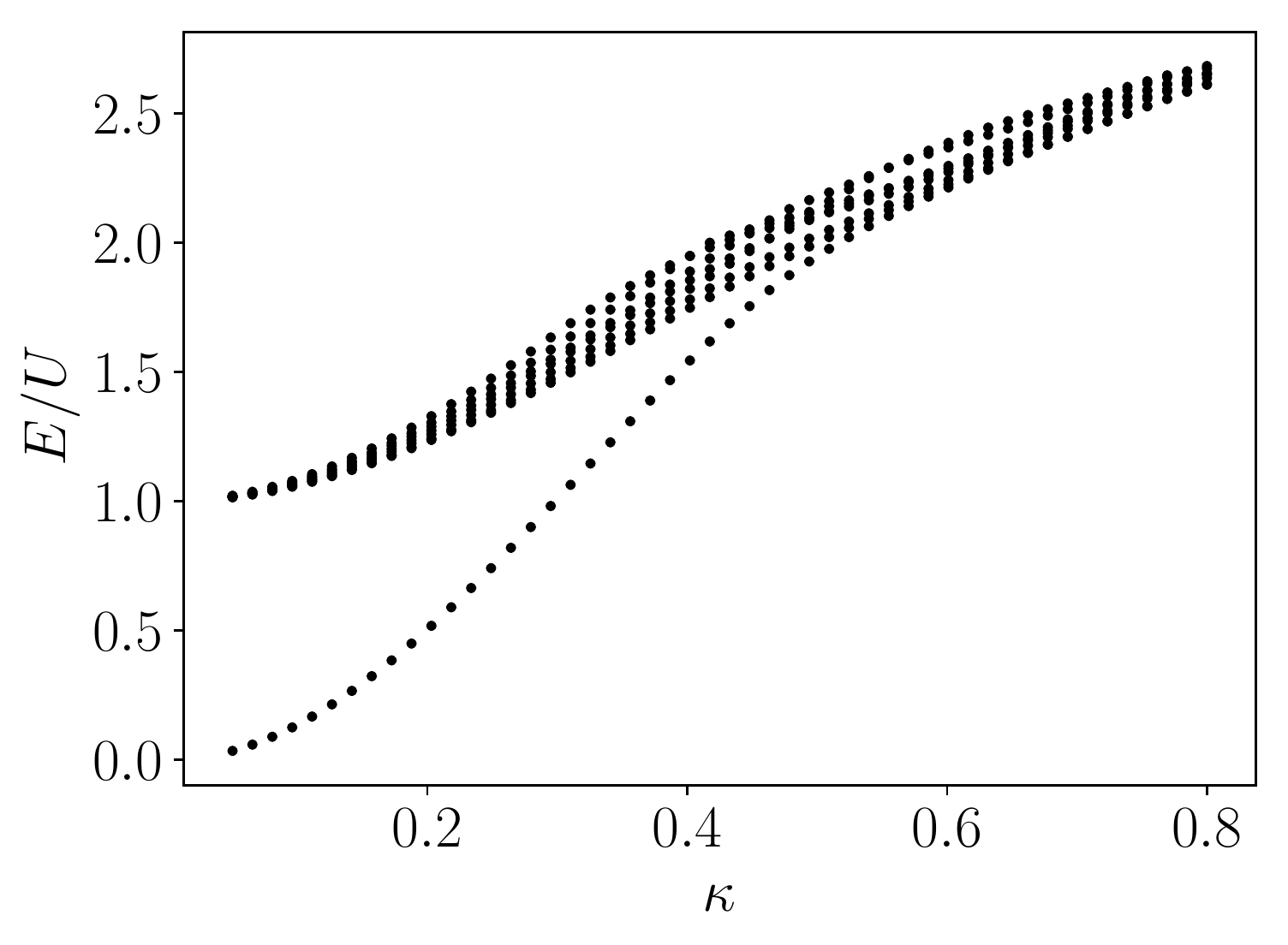}
\caption{Quasienergy spectrum as a function of $\kappa$ for (a) periodic boundary
conditions, and (b) hard walls. Note how in (a) the ground state manifold is always
isolated from the rest of the spectrum, but that in (b) the ground-state merges into
the spectrum as $\kappa$ increases. This makes adiabatic preparation of the cat-state
far more challenging for the case of hard walls.}
\label{Fig:spectrum}
\end{figure}

We show the behaviour of $\chi$ for periodic boundary conditions (the ring) in
Fig. \ref{Fig:cat_contrast_adiabatic_preparation}b, for various ramp speeds.
It can clearly be seen that as the ramp speed decreases, the final value of $\chi$ increases,
indicating that the cat state is being prepared with greater fidelity.
This can be understood from the quasienergy spectrum of the system, shown in Fig. \ref{Fig:spectrum}a.
For small $\kappa$ the system is a Mott insulator, and so the ground state is separated from
the next excited states by a gap of order $U$. Accordingly, as long as the ramp-speed
is sufficiently slow with respect to the gap, $|\dot{\Delta}|/\Delta \ll 1$, the adiabatic
approximation holds and the system safely remains in the ground state. Once
the Mott gap closes, and the system becomes superfluid, the two lowest-lying states
become an almost degenerate doublet. To form the cat correlations, we want the system to remain
in this doublet, without being excited to higher excited states, which again imposes
an adiabatic limit on the speed of the ramp (actually, as noted below, the system barely occupies the first excited state due to symmetry). As the gap is smaller here than in the
Mott regime, the adiabaticity requirement is more stringent, and we can see it is only fulfilled 
when the ramp-time is of the order of $6400 T$.
For more rapid ramps, $\chi$ shows an oscillatory behaviour arising
from the excitation of higher states, which decreases the value of the cat correlations.
To confirm this interpretation, we also measured the overlap (squared) of the final
state with the true ground state of the system for $\kappa = 0.8$. For a ramp-time of
$1600 T$, for example, this takes a value of $0.842$, indicating that a substantial
proportion of the state has been excited out of the ground state. For the slowest ramp,
however, this value rises to $0.999$, demonstrating that the procedure indeed
has excellent fidelity.

In Fig. \ref{Fig:cat_contrast_adiabatic_preparation}c we show the corresponding results
for the hard-wall case. In contrast to the ring, however, even the slowest ramp speeds used are
not able to prepare a cat state with comparable fidelity. As we show in Fig. \ref{Fig:spectrum}b, this
is a consequence of the differences in the energy spectrum produced by the change in boundary
conditions. For hard walls the lowest doublet of states in the superfluid regime is barely
separated from the next-highest states, as we can see explicitly by comparing
Fig. \ref{Fig:low_en_exc_hw} with Fig \ref{Fig:low_exc}. We can again measure the overlap
of the obtained state with the actual ground state of the system, which reveals that even
for a ramp-time of $6400 T$ the fidelity is substantially lower ($\simeq 0.902$) than for the
corresponding case of the
ring, As a result, slower ramp-speeds would
be needed to prepare a cat state, with time scales at least an order of magnitude longer than for the case
of a ring. 

Interestingly, we find that, for both the ring and the hard-wall case, the overlap of the final state with any given excited state is very small. In particular, the overlap with the (also cat-like) first excited state is essentially zero because the ramp-up driving respects the symmetry of the ground state.

\section{Resilience to collapse}
\label{resilience}

Here we focus on the robustness of the cat state against collapse into one of its branches once it has been prepared. By collapse or decay of a cat state we understand its projection into one branch due to energy lowering or information retrieval and lasting for a very long time.

For a boson gas in a ring, one possible cause of collapse is the appearance of a spurious external flux. This can be due to a rotation drift of the optical lattice or to a departure from the ideal switching procedure \cite{Creffield2008} considered in sections \ref{kinetic-driving} or \ref{state-preparation}. 
The spurious flux may be relevant in a SQUID \cite{van2000ch,friedman2000quantum}, where the Josephson coupling generates a periodic flux dependence that can be tuned to form a double-well potential which in turn can be easily imbalanced. Here the physics is different. We have explicitly checked that, under the effect of an external flux, the two condensates shift their average momentum while preserving the cat structure. In the case of a box (hard wall), the situation is even simpler, as any velocity drift can be gauged away.

Particle losses can be a cause of MQS collapse \cite{zoller1998,Glancy2008}. A necessary condition for this to be an efficient decay mechanism is that the emitted atom carries information on the cat branch it comes from. Assuming that upon detection it is possible to know the momentum the atom had in the optical lattice, the information retrieval might be sufficient to cause the collapse. However, this is not clearly the case if we note that the atom with nonzero momentum in the lattice had actually zero group velocity before being emitted (see section \ref{particle-current}). For the reduction cloud, the situation is more radical: even if we were able to know the initial crystal momentum of the detected atom, if this momentum happens to belong to the reduction cloud (i.e. it is different from $\pm \pi/2$), then it carries no information whatsoever on the branch state, since the reduction cloud is identically shared by the two $\pm \pi/2$-momentum condensates.

A similar discussion of atom losses applies to a kinetically-driven boson gas between hard walls.

The interaction with the thermal cloud has sometimes been identified as a possible cause of cat state collapse \cite{moore2006}. Here we have not performed a study of thermal excitations, but the existing experience on conventional boson gases teaches that quasiparticles often reflect the structure of the depletion cloud. The fact that the reduction cloud is common to the two macroscopic condensates strongly suggests that the same is true for the quasiparticles. Thus the assertion that the thermal cloud cannot cause the decay of the MQS state may be viewed as an educated conjecture.

Diagonal impurities may also break the nice degeneracy between cat branches. Both for impurity disorder (see section \ref{state-preparation}) and for isolated impurities, we have numerically checked that the cat structure remains intact within a range of nonzero impurity strength (not shown).

Finally, one may reflect on the role of standard decoherence, i.e., dephasing caused by coupling to an external dissipative environment \cite{von1932mathematische,zurek1982,joos1985emergence,zurek1991physics}. Cold atom systems tend to be isolated and thus insensitive to external sources of dissipation. For the present system, there is an additional, more profound reason to believe that dynamic decoherence is not operating here.
A necessary condition for an environment to cause the collapse of a cat state is that it couples to the system observable whose eigenstates (the ``pointer basis'' of Ref. \cite{zurek1982}) characterize the cat branches. It is difficult to think of a dissipative environment that meets those requirements, considering that the branches of the present state collectively populate nonzero momentum states while yielding individually a zero current average, due to the counterintuitive current operator (see section \ref{particle-current}).

Altogether, there seems to be a number of reasons for asserting that the cat states we have identified in the ground doublet of a kinetically driven boson gas are more protected than most cat states so far investigated. Although we have not been able to check it explicitly, it seems reasonable to venture that this unusual MQS robustness will survive in the thermodynamic limit. Thus the picture emerges of a many-body state with a practically hidden cat structure that only blossoms when subject to the invasive momentum-measurement of a time-of-flight experiment.

\section{Conclusions}

In this paper we have investigated a novel type of Schr\"odinger cat state whose main characteristic is its unusual resilience to collapse into one of its branches.
This is the case of a one-dimensional boson gas (described by the BH model) between hard walls subject to time-periodic driving of the kinetic energy with zero average. As noted in Ref. \cite{pieplow2018generation}, such a boson system system has a preference for collectively populating the states of momentum $\pm \pi/2$. 

We have focused on exploring 
the cat-like nature of the ground state, which has led us to a deeper understanding of this atypical many-body problem. This has been possible thanks to the fortunate fact that, in the ring case and for medium and large driving amplitudes, the time-independent effective Hamiltonian can be approximated by a Hamiltonian which resembles the Richardson-Gaudin pairing model. Our effective system additionally includes an attractive interaction in momentum-space 
\cite{Heimsoth2012} and a peculiar pairing rule that links momenta $k,\pi - k$ (instead of the conventional $k,-k$). This unusual pairing interaction gives rise to a modified depletion cloud which is shared by the two current-carrying condensate branches and which we label reduction cloud to clearly distinguish it from the qualitatively different depletion cloud of the conventional, undriven BH model.

The simplified toy model just described does not apply so well to the boson gas between hard walls. Yet many of its features can still be qualitatively understood in terms of an ensemble of atoms with highly correlated momenta. This picture is quantitatively confirmed by analyzing the many-body ground state in the representation of truncated plane waves (see Fig. \ref{Fig:pw_gr_1exc}). 
We have also studied the case of harmonic confinement and found the same structure of many atoms with a high momentum correlation, which underlines the robustness of the effect.

We have investigated aspects of the ground state such as its intrinsic quality as a cat state, the feasibility of its preparation, and its resistance to collapse. The emerging picture is that of a boson system whose ground doublet is formed by two cat-like states. Of the ground state we can assert that its cat quality and preparation feasibility are acceptable and its resilience to collapse remarkable.

The cat branches differ in the momentum that is macroscopically occupied. Despite their nonzero momentum, each cat branch has separately a zero average current. This counterintuitive result is directly linked to the exotic character of the effective Hamiltonian, which results in an also atypical current operator. The atom momenta can be measured in a time-of-flight experiment, since upon removal of the confining optical lattice the crystal momentum within the lattice becomes the linear momentum in free space. In such an experiment the majority of atoms are expected to flock in a given direction that varies randomly from run to run. In such a scenario one could investigate the statistics of correlations between detected momenta at different times, thus probing also the reduction cloud of the fragmented condensate and its resulting correlations.

In the presence of a finite effective flux, the kinetically driven boson gas displays a nonzero current whose sheer existence depends crucially on the presence of the reduction cloud. This may be viewed as a new form of superfluidity.

The present work paves the way to the study of a novel class of low-energy robust cat states that can be realized in cold atom setups and other similar systems.

\acknowledgments
We would like to thank Anthony Leggett and Peter Zoller for valuable discussions.
This work has been supported by Spain's MINECO through
Grants No. FIS2013-41716-P and FIS2017-84368-P. One of us (FS) would like to acknowledge the 
support of the Real Colegio Complutense at Harvard and the Harvard-MIT 
Center for Ultracold Atoms, where part of this work was done.

\bibliography{catbib}


\appendix

\section{Variational calculation}
\label{app:var}

Take the state 
\begin{align}
|\Psi(\alpha)\rangle &= \alpha|{F} \rangle + \beta | {R}\rangle \\
&
\begin{aligned}
= &\frac{\alpha}{\sqrt{2}}( |N_{-\frac{\pi}{2}},0\rangle+|0,N_{\frac{\pi}{2}}\rangle)  
\\
&+\frac{\beta}{\sqrt{2(L-2)}}(|(N-2)_{-\frac{\pi}{2}},0_{\frac{\pi}{2}}\rangle+|0_{-\frac{\pi}{2}},(N-2)_{\frac{\pi}{2}}\rangle\\
&\times\sum_{k \neq \pm \frac{\pi}{2}} |1_{k}, 1_{\pi-k}\rangle \, ,
\end{aligned}
\label{eq:var_state_1}
\end{align}
where $\alpha, \beta$ are real and  $\alpha^2 + \beta^2 = 1$. 
Here $F$ stands for fragmented condensates and $R$ for reduction cloud.
We find    
\begin{align}
E(\alpha) &= \langle \Psi(\alpha)|\Htm|\Psi(\alpha)\rangle
\\ 
&= \alpha^2(E_{F} - E_{R}) + \alpha\sqrt{1-\alpha^2}\,V + E_{R} \, ,
\end{align}
which is minimized for $\alpha=\alpha_0$ with
\begin{equation}
\alpha_0  = -\text{sgn}(V) \sqrt{\frac{1}{2}+\left(1+\frac{2V^2}{(E_{F}-E_{R})^2}\right)^{-1/2}}
\, .
\label{eq:alpha_min}
\end{equation}
The various constants are 
\begin{align}
&E_{F} = \langle{F}| \Htm| {F}\rangle = -N^2 \, ,\label{eq:EF}\\
&
\begin{aligned}
V& = 2 \langle{F}| \Htm| {R}\rangle \\
&= \sqrt{N(N-1)(L-2)}\, ,
\end{aligned}
\label{eq:V}\\
&\begin{aligned}
E_{R} &= \langle {R} | \Htm |{R}\rangle \\
& = -N^2+4N+L-9 \, .
\end{aligned}
\label{eq:ER}
\end{align}
For unit filling $(N = L)$ the energy reduction for 8 particles on 8 sites due to the cloud is $E(\alpha)-E_{R} = -8.19$ (here, $E_F = -64$, $E(\alpha_0) = -72.2$ and the numerically exact ground state energy is $E_{\rm gs} = -83.2$).  
For $N > 9$  one can prove generally that
\begin{equation}
E(\alpha) < E_{F} ~. 
\end{equation} 
In the limit of large $N$, one obtains
\begin{equation}
E(\alpha_0) - E_F = - N\sqrt{N}\, .
\end{equation}
This clearly shows how the mixing with a cloud of pairs lowers the energy of the fully occupied modes $\pm \pi/2$.  
For simplicity we have omitted the inclusion of $|(N-2)_{-\pi/2},2_{\pi/2}\rangle,|2_{-\pi/2},(N-2)_{\pi/2}\rangle$ because it would have caused the introduction of yet another independent parameter, without further elucidating the effect of interactions of the condensate with its cloud.

In order to better understand the benefit of macroscopically occupying a single mode, we 
may choose a different state as a starting point for the variational calculation. Specifically, we take the ``center" configuration in the expansion of Eq. \eqref{eq:two_mode_gs} and show how it benefits less from the reduction cloud:
\begin{align}
|\Psi_{\rm c} (\alpha)\rangle & =  \alpha|\tilde F\rangle + \beta|\tilde R\rangle \\
&\begin{aligned}
&=\alpha|(N/2)_{-\frac{\pi}{2}},(N/2)_{\frac{\pi}{2}}\rangle\\
&+\frac{\beta}{\sqrt{2(L-2)}}  (|(N/2-2)_{-\frac{\pi}{2}},(N/2)_{\frac{\pi}{2}}\rangle \\
&+ |(N/2)_{-\frac{\pi}{2}},(N/2-2)_{\frac{\pi}{2}}\rangle  )  \sum_{k \neq \pm \frac{\pi}{2}} |1_{k}, 1_{\pi-k}\rangle \, .
\end{aligned}
\label{eq:center_cloud}
\end{align}
When comparing \eqref{eq:center_cloud}  with \eqref{eq:var_state_1}, the key difference is that the two condensate terms in the reduction cloud of \eqref{eq:center_cloud} can mix due to the pair interaction term, namely,
\begin{align}
\begin{aligned}
&\langle (N/2-2)_{-\frac{\pi}{2}},(N/2)_{\frac{\pi}{2}}|h_\infty |(N/2)_{-\frac{\pi}{2}},(N/2-2)_{\frac{\pi}{2}}\rangle \\ 
&=N/2(N/2-1)\neq 0 \, .
\end{aligned}
\label{eq:center_reduct}
\end{align}
By contrast, this mixing does not affect \eqref{eq:var_state_1} because 
their counterparts are not connected by $h_\infty$:
\begin{equation}
\langle (N-2)_{-\frac{\pi}{2}},0_{\frac{\pi}{2}}|h_\infty |0_{-\frac{\pi}{2}},(N-2)_{\frac{\pi}{2}}\rangle = 0~.
\end{equation}
Again we minimize  
\begin{align}
\tilde{E} (\alpha) = \alpha^2(E_{\tilde F} - E_{\tilde R}) + \alpha\sqrt{1-\alpha^2}\,\tilde V + E_{\tilde R} \, ,\label{eq:energy_tilde}
\end{align}%
where 
\begin{align}
&E_{\tilde F} = \langle \tilde F | h_\infty | \tilde F \rangle = - N^2/2 \, , \label{eq:EFt}\\
&\begin{aligned}
\tilde V & = 2 \langle \tilde F | h_\infty | \tilde R\rangle  \\ \nonumber
& = \sqrt{N(N/2-1)(L-2)} \, ,
\end{aligned}
\\ 
&\begin{aligned}
E_{\tilde R} & = \langle \tilde R | h_\infty | \tilde R \rangle \\
& =-N^2/4 + 3N/2 + L - 9  \, .
\end{aligned}\label{eq:ERt}
\end{align}
The value $\tilde \alpha_0$ that minimizes $\tilde{E}(\alpha)$ is formally the same as \eqref{eq:alpha_min} but with Eqs. \eqref{eq:EF}-\eqref{eq:ER} replaced by \eqref{eq:EFt}-\eqref{eq:ERt}. 

Not only is $\tilde E(\tilde \alpha_0) > E(\alpha_0)$ but also the contribution of the reduction cloud to the state and the energy is altered. 
For unit filling $N = L$, the change introduced to $E_{\tilde R}$ due to the mixing in Eq. \eqref{eq:center_reduct} causes $\lim_{L\rightarrow \infty}{\tilde \alpha_0}^2 = 1$ as opposed to $\lim_{L\rightarrow \infty}\alpha_0^2 = 1/2$. 
The closer $\tilde \alpha_0$ is to one, the less important is the contribution of the reduction cloud to $|\Psi_c\rangle$ and to $\tilde E_{\tilde \alpha_0}$, as can be seen in Eq. \eqref{eq:energy_tilde} and the normalization condition for $\alpha,\beta$.  
This means that the reduction cloud in $|\Psi_c(\tilde \alpha_0 )\rangle$ loses importance once the system size increases, which is not the case for $|\Psi(\alpha_0)\rangle$. Even for moderate sizes, such as 8 particles on 8 sites ${\tilde \alpha_0}^2 = 0.92$, whereas $\alpha_0^2 = 0.76$  (the bigger $\alpha$ the smaller the weight of the reduction cloud). We note that \eqref{eq:var_state_1} and \eqref{eq:center_cloud} are just approximate ans\"{a}tze, hence the weight of the condensate does not need to vanish in the thermodynamic limit, as expected for an exact description.

\section{Cat figure of merit}
\label{app:merit} 

We calculate $C$ for one of the states proposed in \cite{dur2002effective} 
\begin{equation}
|\Psi\rangle = \frac{1}{K}(|A\rangle + | B \rangle)
\label{eq:non_ortho_cat}
\end{equation}
where
\begin{align}
&|A\rangle = |N,0\rangle~,\\
&
\begin{aligned}
|B\rangle &= \frac{1}{\sqrt{N!}}(\ce a^\dagger + \se b^\dagger  )^N|{\rm vac} \rangle \\
  &= \frac{1}{\sqrt{N!}} \sum_l \binom{N}{l} \ce^l\se^{N-l}\sqrt{l!(N-l)!}|l,N-l\rangle \, ,
\end{aligned} 
\end{align}
where $\ce\equiv \cos \theta$ and $\se\equiv \sin \theta$.
We find
\begin{align}
C(\theta) &= \sum_k f_k (|\langle A | \xi_k\rangle|-|\langle B | \xi_k\rangle|)^2\\
 &
 \begin{aligned}= 
& \frac{1}{N K^4}\left( \frac{1}{\mathcal{M}_1^2}(N +2N\ce^N + g_1)( N - g_1)^2 
\right. 
 \\
 &\hspace{1cm}\left.+ \frac{g_2^3}{\mathcal{M}_2^2}\right) 
\end{aligned}
\end{align}
\begin{figure} [t!]
	\center
	\includegraphics[width = .9\columnwidth]{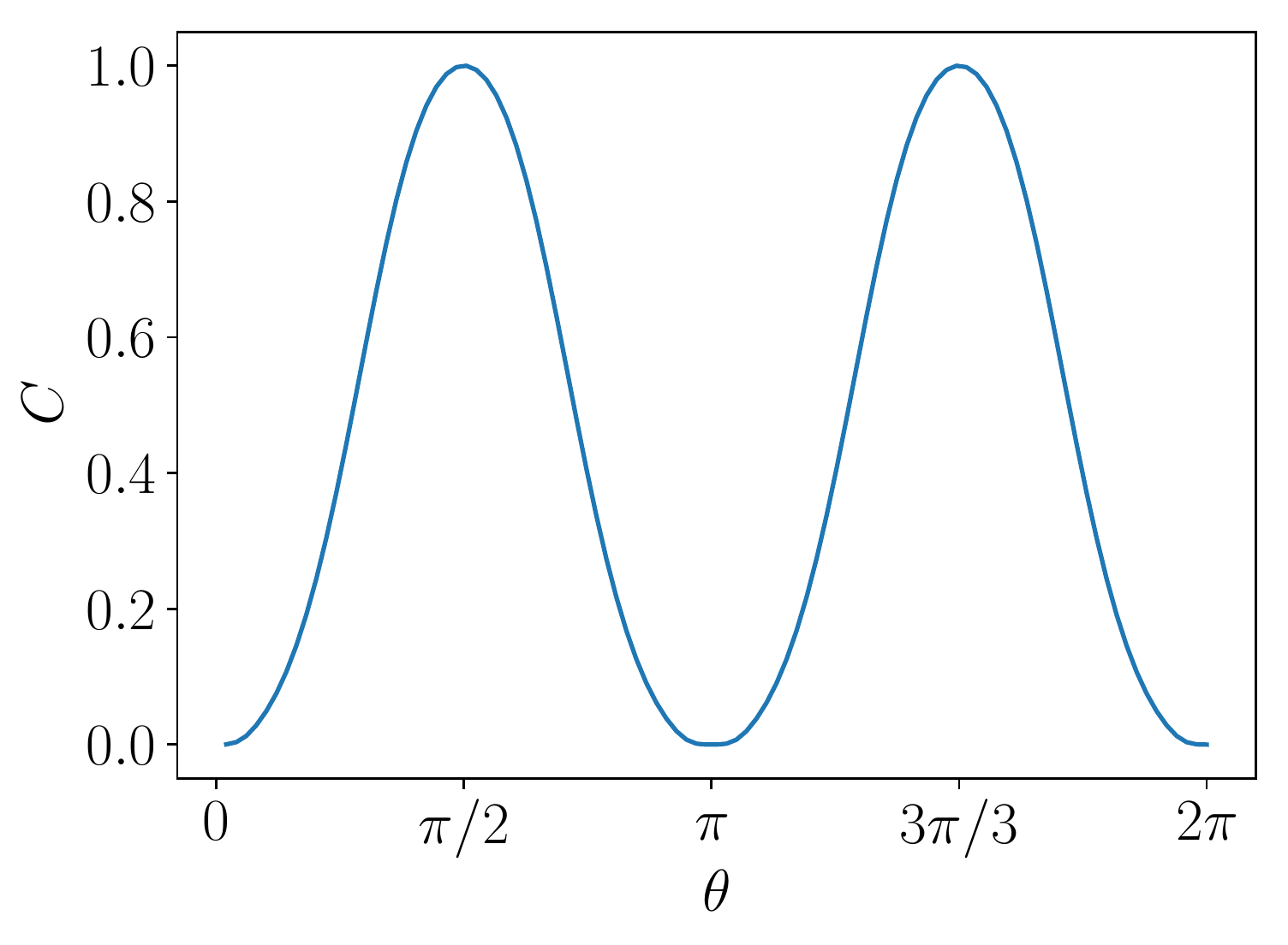}
	\caption{Cattiness $C$ for \eqref{eq:non_ortho_cat}. For $\theta = \pi/2, 3\pi/2$ the state is a perfect cat, for $\theta = 0,\pi$ it has no cat state feature.} 
	\label{Fig:non_ortho_cat}
\end{figure}
where
\begin{align}
&
\begin{aligned}
\mathcal{M}_1^2 = &\frac{1}{K^2}\left( N^2 +2N^2\ce^{N} \vphantom{\frac{1}{N}\sum_a} \right.  + 
\\ 
&\hspace{1cm}\left. \frac{1}{N!}\sum_{l=0}^N\binom{N}{l}^2 \ce^{2l} \se^{2(N-l)}l^2 l!(N-l)! \right)
\end{aligned}
\\
&\mathcal{M}_2^2 = \frac{1}{N!K^2}\sum_{l=0}^N \binom{N}{l}^2 \ce^{2l} \se^{2(N-l)} (N-l)^2l!(N-l)!
\end{align}
and
\begin{align}
&g_1 = \frac{1}{N!}\sum_{l=0}^N \binom{N}{l}^2 \ce^{2l} \se^{2(N-l)}l!(N-l)!l\\
&g_2 = \frac{1}{N!}\sum_{l=0}^N \binom{N}{l}^2 \ce^{2l} \se^{2(N-l)}l!(N-l)!(N-l)
\end{align}
%


\end{document}